%% file: partech-full_arxiv.tex
\documentclass[11pt]{article}
\usepackage{hyperref}
\usepackage{microtype,amsmath,amssymb,latexsym,graphicx, xspace, url, color, subfigure}
\usepackage{amsmath}
\usepackage[margin=1.00in]{geometry}
\usepackage{mathpazo}
\usepackage{calrsfs}
\usepackage{cite}
\usepackage{comment}

\def\reals{{\mathbb R}}
\def\Lspace{{\mathbb L}}
\def\Sspace{{\mathbb S}}

\def\eps{{\varepsilon}}
\def\bd{{\partial}}
\def\A{\mathcal{A}}
\def\B{\mathcal{B}}
\def\C{\mathcal{C}}
\def\F{\mathcal{F}}
\def\G{\mathcal{G}}
\def\K{\mathcal{K}}

\def\R{\mathcal{R}}
\def\S{\mathcal{S}}
\def\T{\mathcal{T}}
\def\U{\mathcal{U}}

\def\VD{\mathrm{VD}}

\def\dist{{\rm dist}}
\def\etal{\textsl{et~al.}}
\def\M{\mathsf{M}}

\def\marrow{\marginpar[\hfill$\longrightarrow$]{$\longleftarrow$}}

\def\micha#1{\textsc{(Micha says: }\marrow\textsf{#1})}
\def\esther#1{\textsc{(Esther says: }\marrow\textsf{#1})}

\newtheorem{theorem}{Theorem}[section]
\newtheorem{lemma}[theorem]{Lemma}
\newtheorem{proposition}[theorem]{Proposition}

\begin{document}


\title{Vertical Decomposition in 3D and 4D with Applications to Line Nearest-Neighbor Searching in 3D
\thanks{%
  Work by Pankaj Agarwal has been partially supported by NSF grants IIS-18-14493, CCF-20-07556, and CCF-22-23870.
  Work by Esther Ezra has been partially supported by Israel Science Foundation Grant 800/22, and also by US-Israel Binational Science Foundation under Grant 2022131.
  Work by Micha Sharir has been partially supported by Israel Science Foundation Grant 260/18.}}

\author{Pankaj K. Agarwal\thanks{
    Department of Computer Science, Duke University, Durham, NC 27708, USA;
    {\sf pankaj@cs.duke.edu,
    https://orcid.org/0000-0002-9439-181X}}
  \and
    Esther Ezra\thanks{
    School of Computer Science, Bar Ilan University, Ramat Gan, Israel;
    {\sf ezraest@cs.biu.ac.il,
      https://orcid.org/0000-0001-8133-1335}}
  \and
    Micha Sharir\thanks{
    School of Computer Science, Tel Aviv University, Tel Aviv, Israel;
    {\sf michas@tauex.tau.ac.il,
      http://orcid.org/0000-0002-2541-3763}}
}

\maketitle


\begin{abstract}
 Vertical decomposition is a widely used general technique for decomposing the cells of arrangements of 
  semi-algebraic sets in $\reals^d$ into constant-complexity subcells. In this paper, we settle in the affirmative a 
  few long-standing open problems involving the vertical decomposition of substructures of 
  arrangements for $d=3,4$: (i) Let $\S$ be a collection of $n$ semi-algebraic sets of constant 
  complexity in $\reals^3$, and let $U(m)$ be an upper bound on the complexity of the union $\U(\S')$ 
  of any subset $\S'\subseteq \S$ of size at most $m$. We prove that the complexity of the vertical 
  decomposition of the complement of $\U(\S)$ is $O^*(n^2+U(n))$ (where the $O^*(\cdot)$ notation 
  hides subpolynomial factors).  We also show that the complexity of the vertical decomposition of 
  the entire arrangement $\A(\S)$ is $O^*(n^2+X)$, where $X$ is the number of vertices in $\A(\S)$.
  (ii) Let $\F$ be a collection of $n$ trivariate functions whose graphs are semi-algebraic sets 
  of constant complexity.  We show that the complexity of the vertical decomposition of the portion of 
  the arrangement $\A(\F)$ in $\reals^4$ lying below the lower envelope of $\F$ is $O^*(n^3)$.  
  
  These results lead to efficient algorithms for a variety of problems 
  involving these decompositions, including algorithms for constructing the decompositions themselves, and for 
  constructing $(1/r)$-cuttings of substructures of arrangements of the kinds considered above. 
  One additional algorithm of interest is for output-sensitive point enclosure queries amid semi-algebraic sets in three or four dimensions.
  
  In addition, as a main domain of applications, we study various proximity problems 
  involving points and lines in $\reals^3$: We first present a linear-size data structure for 
  answering nearest-neighbor queries, with points, amid $n$ lines in $\reals^3$ in $O^*(n^{2/3})$ time 
  per query.
  We also study the converse problem, where we return the nearest neighbor of a query line amid $n$ input 
  points, or lines, in $\reals^3$.
  We obtain a data structure of $O^*(n^4)$ size that answers a nearest-neighbor query in 
  $O(\log n)$ time.
  Finally, We study batched, or offline, variants of these problems, and obtain improved algorithms for 
  such scenarios. 
  
  
\end{abstract}




\section{Introduction}
\label{sec:intro}


Let $\S$ be a family of $n$ semi-algebraic sets\footnote{%
  Roughly speaking, a semi-algebraic set in $\reals^d$ is the set of points in $\reals^d$ that satisfy a Boolean formula over a set of polynomial inequalities; the complexity of a semi-algebraic set is the number of polynomials defining the set and their maximum degree.
  See \cite{BPR} for formal definitions of a semi-algebraic set and its dimension.}
of constant complexity in $\reals^d$. 
The \emph{arrangement} of $\S$, denoted by $\A(\S)$, is the decomposition of $\reals^d$ into 
maximal connected relatively open cells of all dimensions, so that all points within a cell lie in the relative 
interior or boundary of the same subfamily of sets of $\S$. Because of their wide range of applications,
arrangements of semi-algebraic sets have been extensively studied~\cite{AS00,BPR}. The combinatorial
complexity of a cell in $\A(\S)$ can be quite large, and its topology can be quite complex~\cite{AS00}, 
so a fundamental problem in the area of arrangements, for both combinatorial and algorithmic applications,
is to decompose a cell of $\A(\S)$ into constant-complexity subcells, each homeomorphic to 
a ball. In some applications, we wish to decompose all cells of $\A(\S)$ while in others only a subset of cells of $\A(\S)$.

Vertical decomposition is a popular general technique (and perhaps the only general-purpose technique) for constructing such a decomposition.
Roughly speaking, vertical decomposition recurses on the dimension $d$. Let $C$ be a cell of $\A(\S)$.
For $d=2$, the vertical decomposition of $C$ is obtained by erecting a $y$-vertical segment up and down 
from each vertex of $C$ and from each point of vertical tangency on the boundary of $C$,
and extending these segments till they hit another edge of $C$, or else all the way to infinity. This results in a 
decomposition of $C$ into vertical \emph{pseudo-trapezoids} (trapezoids, for short).
For $d=3$, we first erect, upwards and downwards, $z$-vertical \emph{curtains} from each edge of $C$ and
from the silhouette (the locus of points with $z$-vertical tangency) of each $2$-face of $C$, and extend them
until they hit $\bd C$ (or else all the way to infinity). The resulting subcells have a unique pair of
faces as their ``floor'' and ``ceiling,'' but their complexity can still be large. 
In the second decomposition phase, we project each subcell onto the $xy$-plane, apply planar vertical 
decomposition to the projection, and lift each resulting subcell (trapezoid) vertically up to $\reals^3$ 
to the range  between the floor and ceiling of the original subcell. This results in a decomposition of 
$C$ into vertical \emph{pseudo-prisms} (prisms for short), each bounded by up to six facets.
This recursive scheme (on the dimension) can be generalized to higher dimensions, but it
becomes more involved as the dimension grows. In this work, though, we only use the three- and four-dimensional 
scenarios. See~\cite{CEGS-91,Koltun-04a,SA}.

Vertical decompositions, similar to some other geometric decomposition schemes, provide a mechanism for constructing geometric \emph{cuttings} of various substructures of arrangements of semi-algebraic sets~\cite{AS00}, 
which in turn leads to an efficient divide-and-conquer mechanism for solving a variety of combinatorial and algorithmic problems, as well as for constructing data structures for geometric searching problems~\cite{AM94}.
The performance of these algorithms and data structures depends on the
complexity (number of prisms) of the vertical decomposition. For $d=2$, the size of the vertical decomposition of a cell $C$ is proportional to the combinatorial complexity of $C$, but already for $d=3$, the size of the vertical decomposition of $C$ can be $\Omega(n^2)$ even when the complexity of $C$ is $O(n)$. A challenging problem is thus to obtain sharp 
bounds on the complexity of the vertical decomposition of (the cells of) various substructures of $\A(\S)$
for $d\ge 3$.
Despite extensive work on this problem, see, e.g., \cite{AAS-97,AES-99,AS-96,dBGH,CEGS-91,Koltun-04a,SS-97} for a sample of results, several basic problems remain open.
In this paper we settle some of these problems in the affirmative, obtaining sharp bounds on the complexity of the vertical decomposition of various substructures of arrangements, and full arrangements,
for $d=3,4$; see below for a list of our results. As a major application of these results, we study proximity problems involving lines 
and points in $\reals^3$; see below.


\paragraph{Related work.}
Collins~\cite{Col} (see also~\cite{BPR,SS83}) had proposed \emph{cylindrical algebraic decomposition} (CAD) 
as a general technique for decomposing the cells of $\A(\S)$ into pseudo-prisms, in any dimension $d$. However, the number of cells produced is $n^{2^{O(d)}}$. Vertical decomposition can be viewed as an optimized version of CAD, with much smaller complexity. 
Although vertical decompositions for $d=2,3$ have been used since the 1980's~\cite{CI84,CEGSW90}, 
Chazelle~\etal~\cite{CEGS-91} described the construction of vertical decomposition in general,
for arrangements of semi-algebraic sets in $\reals^d$, and proved a bound of $O^*(n^{2d-3})$ for $d\ge 3$
(where the $O^*(\cdot)$ notation hides subpolynomial factors). They also showed that the vertical decomposition of $\A(\S)$ can be computed in $O^*(n^{2d-3})$ expected time.
The bound was improved to $O^*(n^{2d-4})$, for $d \ge 4$, by Koltun~\cite{Koltun-04a}.
These bounds are nearly optimal for $d\le 4$, and are strongly suspected to be far from optimal for $d\ge 5$. Improving the bound, for $d\ge 5$, is a major 30-years-old open problem in this area (which we do not address in this work).

In many applications, one is interested in computing the vertical decomposition of (the cells of) only a substructure of $\A(\S)$.
In this case, the goal is to show that if the substructure under
consideration has asymptotic complexity $o(n^d)$, then so should be the complexity of
its vertical decomposition. This statement is true in the plane, as already mentioned, and has been shown to
hold for arrangements of triangles in 3D~\cite{dBGH,Tag}.
Notwithstanding a few results on the vertical decompositions of substructures of 3D and 4D arrangements, 
see, e.g.,~\cite{AAS-97,AES-99,AS-96,SS-97}, the aforementioned fundamental problem has remained 
largely open for $d\ge 3$.
For example, even though the complexity of the union of a set of objects in $\reals^3$ in many interesting cases---such as a set of cylinders or a set of fat objects---is known to be 
$O^*(n^2)$~\cite{AS-00, AEKS-06,Ezra-11,ES-09}, 
no subcubic bound was known on the size of the vertical decomposition of the complement of their union.
In $\reals^4$, the complexity of the lower envelope of $n$ trivariate functions (whose graphs are semi-algebraic sets of constant complexity) is $O^*(n^3)$
(see, e.g.,~\cite{SA}), however, no $o(n^4)$ bound was known on the complexity of the corresponding vertical decomposition
of the minimization diagram, which is the $xyz$-projection of the lower envelope.

We conclude this discussion by noting that special-purpose decomposition schemes have been proposed
for decomposing cells in arrangements of hyperplanes, boxes, or simplices, using triangulations, 
binary space partitions, or variants of vertical decomposition; see, e.g., \cite{AS00,ASS21,AS-90,HS96} and references therein. Some of these methods also work for arrangements of semi-algebraic sets using the so called linearization technique~\cite{AM94}, albeit yielding in general much weaker bounds.

\paragraph{Our contributions.}
The paper contains three sets of main results --- (i) sharp bounds on the complexity of 
vertical decompositions 
of substructures of arrangements in $\reals^3$ and $\reals^4$, (ii) efficient algorithms for 
constructing these decompositions and related structures, and (iii) as a major application domain,
efficient data structures for line-point proximity problems in $\reals^3$.

\smallskip
\noindent\textbf{\textit{Vertical decomposition.}} We make significant progress on bounding the size of the
vertical decomposition of substructures of arrangements in $\reals^3$ and $\reals^4$, 
by establishing the following combinatorial bounds.

\smallskip
\textit{Union of semi-algebraic sets.}
Let $\S$ be a family of $n$ semi-algebraic sets of constant complexity in $\reals^3$, 
and let $U(m)$ be an upper bound on the complexity of the union $\U(\S')$ of any subset $\S' \subseteq \S$ of size at most $m$, for any $m>0$. (Note that, by definition, $U(m)$ is monotone increasing in $m$.) We show that the complexity of the vertical decomposition of the complement of 
$\U(\S)$ is $O^*(n^2 + U(n))$ (Section~\ref{sec:union}).

\smallskip
\textit{Lower envelopes.}
Let $\F$ be a collection of $n$ trivariate functions whose graphs are semi-algebraic sets of constant 
complexity, and let $\A(\F)$ denote the arrangement (in $\reals^4$) of their graphs. 
The \emph{lower envelope} $E_\F$ of $\F$ is defined as 
$E_\F(x) = \min_{F\in\F} F(x)$, for $x\in\reals^3$.
We show that the complexity of the vertical decomposition of the cell of\footnote{%
  Even though this vertical decomposition is in $\reals^4$, it is effectively obtained from the vertical decomposition of the \emph{minimization diagram} of $E_\F$ in $\reals^3$; see below for details.}
$\A(\F)$ lying  below (the graph of) $E_\F$ is $O^*(n^3)$, 
thereby matching the general upper bound on the complexity of lower envelopes in $\reals^4$~\cite{SA}
(Section~\ref{sec:env}).

\smallskip
\textit{Sparse arrangements.}
Let $\S$ be a collection of $n$ semi-algebraic sets of constant complexity in $\reals^3$, and let $X$ denote 
the number of vertices in $\A(\S)$. 
We show that the complexity of the vertical decomposition of the entire arrangement $\A(\S)$ is $O^*(n^2 + X)$ 
(Section~\ref{sec:output}).

\smallskip
\noindent\textbf{\textit{Algorithms.}}
There are a few immediate algorithmic consequences of our combinatorial results:

\smallskip
\textit{Computing vertical decompositions.}
All these vertical decompositions can be constructed, namely, the set of pseudo-prisms in the vertical decomposition can be computed, in time comparable with their respective complexity bounds. 
Section~\ref{subsec:alg} describes the construction for the complement of the union of semi-algebraic sets 
in $\reals^3$, as well as for the lower envelopes (or rather minimization diagrams) of trivariate functions 
(whose graphs are semi-algebraic sets of constant complexity); the same approach extends to sparse arrangements. 
We note that Agarwal~\etal~\cite{AAS-97} described a randomized algorithm for constructing the vertices, edges, and $2$-faces of the minimization diagram of a set of trivariate (constant-complexity semi-algebraic) 
functions in $O^*(n^3)$ expected time. In addition, with $O^*(n^3)$ preprocessing, their technique 
can also compute, in $O(\log n)$ time, the function that appears on the lower envelope for a query point $\xi\in\reals^3$. 
(Their algorithm can also compute, in $O^*(n^2+U(n))$ expected time,
the vertices, edges, and $2$-faces of the union of a collection $\S$ of semi-algebraic sets in $\reals^3$, 
where $U(m)$, as above, is the maximum complexity of the union of a subset of $\S$ of size $m$.)
However, their algorithm does not compute three-dimensional cells of the minimization diagram, nor does 
it compute the vertical decomposition of the minimization diagram. See also~\cite{AS-96}.

\textit{Geometric cuttings.}
Let $\S$ be a collection of $n$ semi-algebraic sets of constant complexity in $\reals^d$.
Let $\Pi$ be a substructure of $\A(\S)$, defined by a collection of cells of $\A(\S)$ that 
satisfy certain properties (e.g., lying in the complement of the union or lying below the lower envelope).
For a parameter $r>1$, a \emph{$(1/r)$-cutting} of $\Pi$ (with respect to $\S$) is a set 
$\Xi$ of pseudo-prisms with pairwise-disjoint relative interiors that cover $\Pi$ such that the
relative interior of each pseudo-prism $\tau \in \Xi$ is crossed by (intersected by but not contained in) at 
most $n/r$ sets of $\S$. The subset of $\S$ crossed by $\tau$ is called the \emph{conflict list} of $\tau$.
Our combinatorial results lead to the construction of small-size $(1/r)$-cuttings of $\Pi$. Their
size is dictated by our new bounds for the complexity of the vertical decomposition of $\Pi$. 
For the case of the complement of the union of sets in $\reals^3$, the bound is $O^*(r^2 + U(r))$.
For the case of the region below the lower envelope of trivariate functions in $\reals^4$, the bound is 
$O^*(r^3)$.
For the case of an entire three-dimensional arrangement of complexity $X$, 
we obtain a $(1/r)$-cutting of $\A(\S)$, for any parameter $r\le n$, of total complexity 
$O^*(r^2 + r^3X/n^3)$. 
The cuttings along with the conflict lists of all of its cells can be constructed in $O(n)$ expected time if 
$r$ is a constant (Section~\ref{sec:cutting}).

\smallskip
\textit{Point-enclosure queries.}
Let $\S$ be a family of $n$ semi-algebraic sets in $\reals^3$, and let $U(\cdot)$ denote a bound on its union complexity, as above.
We obtain a data structure of size and preprocessing cost $O^*(n^2 + U(n))$ that, for a query point $q\in\reals^3$, returns all $k$ sets of $\S$ containing $q$ in $O^*(1 + k)$ time.
Similarly, for a given family $\F$ of $n$ semi-algebraic trivariate functions, we can construct a data structure of size $O^*(n^3)$ that, for a query point $q \in \reals^4$, can report, in $O^*(1 + k)$ time, all the $k$ functions of $\F$ whose graphs lie below $q$.

\smallskip
\noindent\textbf{\textit{Proximity problems for points and lines in $\reals^3$.}}
In the third part, building on our vertical-decomposition and geometric-cutting results,
we present efficient data structures and algorithms for various proximity problems 
involving points and lines in $\reals^3$.  

\smallskip
\textit{Nearest line-neighbor to a query point.}
A set $L$ of $n$ lines in $\reals^3$ can be preprocessed, in $O(n\log n)$ expected time,
into a data structure of size $O(n)$, so that for a query point $q\in\reals^3$, 
the nearest neighbor of $q$ in $L$  can be returned in $O^*(n^{2/3})$ time (Section~\ref{sec:plnn}).
We note that a linear-size data structure with $O^*(n^{3/4})$ query time can be obtained by mapping each 
line of $L$ to a point in $\reals^4$ and using four-dimensional semi-algebraic range 
searching techniques~\cite{AMS}.
We also note that a data structure of $O^*(n^3)$ size and $O(\log n)$ query time can be obtained by constructing and preprocessing the Voronoi diagram of the lines in $L$ for point-location queries, 
following an approach similar to that in~\cite{MS}.

Our data structure constructs a partition tree, as in~\cite{AM94,ShSh}, using geometric cuttings. The main challenge in adapting these preceding approaches to our setting is the construction of a so-called \emph{test set}, namely, a small set of representative queries (typically more involved than the usual queries) so that if the data structure can answer those queries efficiently 
then it can answer efficiently the query for any point in $\reals^3$. Our new results on vertical decomposition of the lower envelope of trivariate functions and on geometric cuttings provide the missing ingredients needed for constructing such test sets. See Section~\ref{sec:plnn} for details.


\smallskip
\textit{Nearest point-neighbor to a query line.}
We can preprocess a set $P$ of $n$ points in $\reals^3$, in expected $O^*(n^4)$ time, into a data structure 
of $O^*(n^4)$ size, so that, 
for a query line $\ell$ in $\reals^3$, its nearest neighbor in $P$ can be returned
in $O(\log n)$ time (Section~\ref{sec:Fast_NN1}). 
The standard tools would yield a data structure of size $O^*(n^5)$ for answering fast queries.

Roughly speaking, after applying some geometric transformations, we reduce the nearest-neighbor query to 
a point-location query in a sandwich region enclosed between two envelopes of trivariate functions. 
As we do not know how to perform this task efficiently in a direct manner, due to the lack of a good bound on the complexity of the vertical decomposition of such a region (see~\cite{KS}, where this is stated as a major open problem), we use a more involved scheme that achieves the desired efficiency.

We note that a linear-size data structure with $O^*(n^{2/3})$ query time can be obtained by using 
known results on 3D semi-algebraic range searching~\cite{AMS}.
Our new results on vertical decomposition of the complement of the union of objects 
in $\reals^3$ leads to a faster solution to a restricted version of this problem. That is, 
we can preprocess a set of $n$ points in $\reals^3$ into a linear-size data structure that returns,
in $O^*(n^{1/2})$ time, a point within distance at most $1$ from a query line, if there exists one. 
This problem was recently studied in Agarwal and Ezra~\cite{AE-23}, and they had 
obtained a more involved data structure with a similar bound. By combining our vertical-decomposition result with 
some of their ideas, we obtain a significantly simpler data structure.

\smallskip
\textit{Nearest line-neighbor to a query line.}
We can preprocess a set $L$ of $n$ lines in $\reals^3$, in $O^*(n^4)$ expected time,
into a data structure of size $O^*(n^4)$, so that the nearest neighbor in $L$ of a query line can be computed in $O(\log n)$ time (Section~\ref{sec:Fast_NN2}).
Again, we note that a linear-size data structure with $O^*(n^{3/4})$ query time can be obtained by
using standard four-dimensional semi-algebraic range searching techniques~\cite{AMS}, and that a structure of size $O^*(n^5)$ for the fast query regime can also be obtained by standard methods.

\smallskip
\textit{Off-line nearest-neighbor queries.}
Let us now consider the case when all queries are given in advance. That is, we have a set $L$ of $n$ lines 
and a set $P$ of $m$ points in $\reals^3$, and the goal is to compute the nearest neighbor in $P$ of each line of $L$.
We present a randomized algorithm with $O^*(m^{4/7}n^{6/7} + m + n)$ expected running time (Section~\ref{sec:ann}). We note that by plugging our on-line algorithm with the standard space/query-time trade-off techniques would lead to an algorithm with $O^*(m^{8/11}n^{9/11}+m+n)$ expected running time.




\section{Vertical Decomposition of the Complement of the Union}
\label{sec:union}


Let $\S$ be a collection of $n$ semi-algebraic sets of constant complexity in $\reals^3$.
For any subset $\S'$ of $\S$, let $\U(\S')$ denote the union of $\S'$, and let $\C(\S')$ denote 
the complement of $\U(\S')$. Let $U(m)$ denote the maximum complexity of $\U(S')$---namely, the 
number of vertices, edges 
and $2$-faces of the union boundary---over all subsets $\S'$ of size at most $m$. Clearly  $U(m) = O(m^3)$, 
but as mentioned in the introduction, 
$U(m) = O^*(m^2)$ in many interesting cases.
Let $\VD(\S)$ denote the \emph{vertical decomposition} of $\C = \C(\S)$, and let $C(n)$ denote the 
maximum complexity of $\VD(\S)$, where the maximum is taken over all collections of $n$
semi-algebraic sets of constant complexity. Our goal is to obtain a sharp bound on $C(n)$.

A pair $(e,e')$ of edges of $\A(\S)$ is called \emph{vertically visible} if there exists a vertical line 
$\lambda$ that meets both $e$ and $e'$, so that the relative interior of the segment of $\lambda$ 
connecting $e$ and $e'$ does not meet the boundary of any set of $\S$, and we refer to 
 the pair  of points $(\lambda\cap e, \lambda \cap e')$ as a \emph{vertical visibility}. 
A pair $(e,e')$ of edges can give rise to more than one but at most $O(1)$ vertical visibilities.
It is well known (see, e.g., \cite{SA}) that
$C(n)$ is proportional to $U(n)$ plus the number of vertical visibilities between pairs of 
edges of $\bd\U$ that occur within $\C$, so it suffices to bound the latter quantity.

To bound the number of vertical visibilities, we fix an edge $e$ of $\bd\U$, regarding $e$ as the lower
edge in the vertical visibilities that we seek,\footnote{%
  We assume that the two sets whose boundaries intersect at $e$ lie locally below $e$, for otherwise
  $e$ cannot play the role of the bottom edge of a vertically visible pair in the complement of the union.}
and erect a \emph{vertical curtain} $V(e)$ over $e$, 
which is the (two-dimensional) union of all $z$-vertical rays emanating upwards from the points of $e$.
The boundary of each set $S\in\S$ (ignoring the two that form $e$) intersects $V(e)$ in a 
one-dimensional curve $\gamma_S$, which can be empty or disconnected, but is of constant complexity.
Note that none of the curves $\gamma_S$ cross $e$, for such an intersection would be a vertex of
the arrangement of $\S$ and, by definition, $e$ cannot contain such a vertex.

We form the lower envelope $E_e$ of the curves $\gamma_S$, and note that each breakpoint $a$ of $E_e$,
at which two curves meet, lies on some edge $e'$ of $\bd \U$ which forms a vertically visible pair 
with $e$, with the vertical visibility taking place between $a$ and $e$. 
The other breakpoints, formed at endpoints of connected portions of the curves, occur 
when a vertical line (supporting a ray of the curtain $V(e)$) is tangent to some $S\in\S$; 
that is, the breakpoint occurs on the vertical silhouette of $S$. It is easy to show that the overall 
number of vertical visibilities involving silhouettes is only $O^*(n^2)$. Indeed, there are $O(n)$ silhouettes, 
each of constant complexity, and the vertical visibilities that they are involved in correspond to 
breakpoints of lower or upper envelopes within the vertical curtains that they span. As each envelope
can be regarded as the lower envelope of univariate functions, it has $O^*(n)$ complexity~\cite{SA},
and the claim follows.

To facilitate the forthcoming analysis, we turn the problem into a bipartite problem, where each
set of $\S$ is assigned at random a color red or blue, yielding a partition $\S = \R\cup\B$,
where $\R$ (resp., $\B$) is the set of all red (resp., blue) sets, and our goal is to
bound the number of vertical visibilities between red-red edges (edges formed by the intersection
of the boundaries of two red sets) and blue-blue edges (those formed by the intersection
of the boundaries of two blue sets).
Note that a red-red edge $e$ on the boundary of the union of $\R$ is not necessarily an original 
edge of the boundary of $\U(\S)$, as $e$ may contain red-red-blue vertices (or even be fully contained 
in a blue set). Still, if there exists a vertical visibility in $\C(\S)$ whose lower endpoint
$b$ lies on $e$, then $b$ lies on a portion of $e$ that forms an edge of $\bd\U(\S)$. 
Of course, not all vertically visible pairs are captured in 
this coloring scheme. Nevertheless, it is easily checked that the expected number of visible 
pairs with this coloring is $1/8$ of the overall number of visible pairs, so, up to this factor, 
there is no loss of generality in using this coloring scheme.

So the setup that we face is: We are given a set $\R$ of $m$ red sets and a set $\B$ of $n$ blue 
sets (in the above scheme, both $m$ and $n$ are half the size of $\S$ in expectation), and our 
goal is to bound the number $C(m,n)$ of vertical visibilities between pairs $(e,e')$ of edges, 
where $e$ is a red-red edge and $e'$ is a blue-blue edge, and the vertical visibility takes place
in the complement of $\U(\R\cup\B)$.

We estimate $C(m,n)$ using an extension of the recursive analysis in \cite[Section 2]{KS}.\footnote{%
  We credit this work for providing us the initial inspiration that their technique can be adapted to apply in our settings too.}
We fix some sufficiently large constant parameter $k$, and partition $\B$ arbitrarily into $k$ 
subsets $\B_1,\ldots,\B_k$, each of size $n/k$ (ignoring rounding issues). We solve the problem 
recursively for $\R$ and 
each $\B_i$. Each subproblem yields at most $C(m,n/k)$ vertical visibilities. Note that these
vertical visibilities are not necessarily vertical visibilities in the full red-blue setup, 
because sets in other subsets $\B_j$ may show up between the edges in such a pair and 
destroy the vertical visibility between them. Nevertheless, each original vertical
visibility is either one of these recursively obtained visibilities, or arises at a pair 
$(e,e')$ where $e$ is a red-red edge and $e'$ is a blue-blue edge formed by the intersection 
of two boundaries of sets in different subsets $\B_i$, $\B_j$. 
We now proceed to bound the number of pairs of the latter kind.

To do so, fix a red-red edge $e$, and assume that $e$ plays the role of the bottom edge in a
vertically visible pair. Consider the upward vertical curtain $V(e)$ of $e$, and form within $V(e)$ 
the $k$ blue envelopes $E_e^{(1)},\ldots,E_e^{(k)}$, where $E_e^{(i)}$ is the lower envelope of the 
curves $\gamma_S$, for $S\in\B_i$, for $i=1,\ldots,k$. The breakpoints of the envelopes (ignoring
silhouette breakpoints) correspond to recursively obtained pairs $(e,e')$ (as noted, not all breakpoints yield
visibilities in the full setup), but we are also interested 
in the additional breakpoints of the 
overall lower envelope $E_e$ of these $k$ envelopes.

Let $M_e^{(i)}$ denote the number of breakpoints of $E_e^{(i)}$, for $i=1,\ldots,k$, and put
$M_e = \sum_i M_e^{(i)}$. Notice that $\sum_e M_e^{(i)}$ is the number of vertical visibilities
between $\R$ and $\B_i$, so it is at most $C(m,n/k)$. Thus $\sum_e M_e \le kC(m,n/k)$.

Inspired by the analysis in~\cite{KS},
we follow a technique similar to one used by Har-Peled~\cite{HP-99} in a different context. Specifically, we partition $V(e)$ into vertical sub-curtains $V_1(e),\ldots,V_t(e)$ by upward vertical rays, 
so that the overall number of breakpoints of the individual envelopes within each sub-curtain
is $k$, except possibly for the last sub-curtain, where the number is at most $k$, so $t \le 1 + M_e/k$. 
Within each sub-curtain $V_j(e)$ there are only at most $2k$ 
blue curves $\gamma_S$ that participate in the envelopes $E_e^{(i)}$, of which $k$ show up on 
the envelopes at an extreme ray of $V_j(e)$, and at most $k$ others replace them along the various envelopes, within the 
sub-curtain. Hence, within any fixed $V_j(e)$, $E_e$ is the lower envelope of at most $2k$ connected
subarcs of boundary 
curves $\gamma_S$, so its combinatorial complexity is at most $\lambda_s(2k)$, where
$\lambda_s(m)$ is the near-linear maximum length of Davenport-Schinzel sequences of order $s$ 
on $m$ symbols, for some constant parameter $s$ that depends on the complexity of the sets 
of $\S$~\cite{SA}. We write this bound as $k\beta(k)$, for an appropriate near-constant
extremely slowly growing function $\beta(k)$, and conclude that the number of breakpoints 
of $E_e$ within each sub-curtain is at most $k\beta(k)$, for a total of at most
$kt\beta(k) = (k+M_e)\beta(k)$ breakpoints. Summing over all red-red edges $e$, 
we obtain 
\[
C(m,n) \le \left(\sum_e (k+M_e)\right)\beta(k) \le k\beta(k) C(m,n/k) + k\beta(k) U(m) .
\]

We next switch the roles of red and blue, and apply the same analysis to each pair $\R$, $\B_i$
of sets, keeping $\B_i$ fixed and partitioning $\R$ into $k$ subsets of size $m/k$ each. 
(We now reverse the direction of the $z$-axis, considering downward-directed vertical 
curtains erected from the edges formed by the sets of $\B_i$.)
The analysis proceeds more or less verbatim, and yields the following bound on the number of vertical visibilities:
\[
	C(m,n) \le k^2\beta^2(k) C(m/k,n/k) + k^2\beta^2(k) U(n/k) + k\beta(k) U(m) 
\]
If $U(m) = O^*(m^2)$, we obtain the recurrence
\[
C(m,n) \le k^2\beta^2(k) C(m/k,n/k) + k\beta(k) O^*(m^2) + \beta^2(k) O^*(n^2) .
\]
Note that the right-hand side of this recurrence also subsumes the number of $O^*(m^2+n^2)$
vertical visibilities that involve the silhouettes of the red and blue sets. 

We solve this recurrence for the original setup, where $m$ and $n$ are both roughly half the
total number of sets, which we continue to denote by $n$, with some abuse of notation.
By choosing $k$ to be a sufficiently large constant, the solution of the resulting recurrence
is $O^*(n^2)$. We thus conclude that the number of vertical visibilities between pairs of 
edges of $\U(\S)$ is $O^*(n^2)$.
A similar analysis applies when $U(n)$ is superquadratic. In this case the bound on the complexity of
the vertical decomposition is $O^*(U(n))$, as is easily checked. Putting everything together, we obtain the following main result of this section.

\begin{theorem}
  \label{thm:main}
  Let $\S$ be a collection of $n$ constant-complexity semi-algebraic sets in $\reals^3$, with an upper bound $U(m)$ 
  on the combinatorial complexity of the union of any subset of $\S$ of size $m$.
  Then the size of the vertical decomposition of the complement of the union of $\S$
  is $O^*(n^2+U(n))$.
\end{theorem}


\section{Vertical Decomposition of Lower Envelopes in $\reals^4$}
\label{sec:env}



Let $\F$ be a collection of $n$ trivariate semi-algebraic functions of constant complexity,
let $E = E_\F$ denote the lower envelope of $\F$, let $E^- = E^-_\F$ denote the portion of $\reals^4$ below $E$, 
and let $M = M_\F$ denote the minimization diagram of $E$, namely the projection of $E$ onto the
$xyz$-space. Our goal is to estimate the combinatorial complexity of the vertical decomposition of $M$.
This three-dimensional decomposition can then be lifted up in the $w$-direction to induce a suitable
decomposition of $E^-$, which we refer to as the vertical decomposition of $E$.
We note that the complexity of (the undecomposed) $E$ and of $M$ is $O^*(n^3)$~\cite{SA}.
The main result of this section yields the same asymptotic bound for their vertical decomposition:
\begin{theorem} \label{thm:env4d}
The complexity of the vertical decomposition of the lower envelope (that is, of the minimization diagram) 
of a collection of $n$ trivariate semi-algebraic functions of constant complexity is $O^*(n^3)$.
\end{theorem}

\noindent{\bf Proof.}
We assume that the functions of $\F$ are in general position, continuous and totally defined.
None of these assumptions are essential, but they simplify the analysis. 
We identify each function of $\F$ with its three-dimensional graph. 
We recall the way in which the vertical decomposition $\VD$ of $M$ is constructed. 
We fix a function $a$ in $\F$. For each function $b\in\F\setminus\{a\}$, 
we use $\sigma_{ab} = \sigma_{ba}$ to denote the $xyz$-projection of the two-dimensional intersection 
surface $a\cap b$. The surface $\sigma_{ab}$ partitions the $xyz$-space into the regions $\sigma_{ab}^+$ and
$\sigma_{ab}^-$, where $\sigma_{ab}^+$ (resp., $\sigma_{ab}^-$) consists or those points $(x,y,z)$
for which $a(x,y,z) \ge b(x,y,z)$ (resp., $a(x,y,z) \le b(x,y,z)$). We observe that the complement
$\C_a$ of the union $\U_a := \bigcup \left\{ \sigma_{ab}^+ \mid b \in\F\setminus\{a\} \right\}$ 
is precisely the portion of the $xyz$-space over which $a$ attains the envelope $E$.

We now construct the three-dimensional vertical decomposition, denoted as $\VD_a$, of $\C_a$,
and repeat this construction to each complement $\C_a$, over $a\in\F$, observing that the regions
$\C_a$ are pairwise openly disjoint. The union of all these decompositions yields the vertical
decomposition of $M_\F$, and, as mentioned above, the vertical decomposition of $E_\F$ is obtained 
by lifting this decomposition to $E_\F$ (or to $E_\F^-$, see below), in a straightforward manner.

We comment that, as already noted, we can also obtain by this approach the vertical decomposition of $E^-$. Each cell $\tau$ in the decomposition of $M$ is lifted to the semi-unbounded region
\[
\{ (x,y,z,w) \mid (x,y,z)\in\tau \;\text{and}\; w\le E(x,y,z) \} .
\]

We have thus (almost) reduced the problem to that studied in Section~\ref{sec:union}.
The difference is that there we assumed that the complexity of the union of any subcollection of at most $m$ of the given objects is $O^*(m^2)$, or at least that we have some (subcubic) bound $U(m)$ on that complexity.
Here, though, this no longer holds. That is, considering 
the entire collection $\F$, and denoting by $M_a$ the complexity of $\U_a$, all we know is that 
$\sum_a M_a = O^*(n^3)$, so we have the bound $O^*(n^2)$ only for the \emph{average} value of $M_a$. 
To overcome this technicality, we modify the previous analysis as follows.

Recall that in Section~\ref{sec:union} we have reduced the problem to a bichromatic problem by
assigning to each object the color red or blue at random. Here we extend this technique to obtain
a trichromatic reduction, by assigning to each function the color red, blue or green at random. 
We now consider only unions $\U_a$ for green functions $a$, and within the complement $\C_a$ of
any of these unions, we only consider vertical visibilities between red-red edges and blue-blue
edges (technically, they are green-red-red and green-blue-blue edges), exactly as in 
Section~\ref{sec:union}. Again, any vertical visibility that arises in the original 
decomposition has a constant probability to show up as a green-red-red vs.~green-blue-blue 
visibility in the trichromatic version.

For each green function $a$, the overhead terms that appear in the analysis can be written as
$M(\{a\},\R,\B)$ and $M(\{a\},\R,\B_i)$, where, for arbitrary sets $\G$, $\R$, $\B$ of green, red, and blue objects, respectively, $M(\G,\R,\B)$
denotes the number of the green-red-red and green-blue-blue edges of the undecomposed envelope of
$\G\cup\R\cup\B$. 
Here $\R$, $\B$, and the $\B_i$'s may be recursively obtained subsets of the original sets. Summing these
quantities over $a$, 
we obtain $M(\G,\R,\B)$ and $M(\G,\R,\B_i)$, respectively.
We also use the notation $M(u,v,w)$ to denote the maximum value of
$M(\G,\R,\B)$ for $|\G| \le u$, $|\R| \le v$ and $|\B| \le w$.

Consider, say, a green-red-red edge $e$ that appears on the boundary of (the complement $\C_a$ of) the union 
$\U_a$ for some green function $a$ (the same argument holds for green-blue-blue edges). 
If we replace $\G$ by a subset $\G'$ that contains $a$, $\C_a$ can only grow, 
since fewer regions $\sigma_{ab}^+$ form the union $\U_a$. Hence $e$ does not disappear,
and can only extend, possibly even merge with other edges formed by the same triple of functions.
In particular, the number of vertical visibilities in $\C_a$ between green-red-red edges and
green-blue-blue edges can only increase.

We use this observation as follows. In the first two-step recursive round, as described in
Section~\ref{sec:union}, we first partition $\G$ into $k$ subsets $\G_1,\G_2,\ldots,\G_k$, each of 
size $n/k$, apply the analysis to each $\G_i$ and $\R$ and $\B$, and then sum up the resulting bounds
for $i=1,\ldots,k$. Denote by $C(u,v,w)$
the maximum number of vertical visibilities for sets of at most $u$ green, $v$ red, and $w$ blue functions. 
The overhead term will be at most $O(M(u,v,w)) = O^*((u+v+w)^3)$, and the recursive term
will be at most $C(u/k,v/k,w/k)$ at each recursive subproblem.
Therefore, 
by applying the recursive relation from Section~\ref{sec:union} on the number of red-blue vertical visibilities, we obtain the recurrence:
\[
C(u,v,w) \le \sum_{i=1}^k k^2\beta^2(k) C(u/k,v/k,w/k) + k\beta(k) M(u/k,v,w) ,
\]
which leads to the recursive relation:
\[
C(u,v,w) \le k^3\beta^2(k) C(u/k,v/k,w/k) + k^2\beta(k) O^*((u+v+w)^3) .
\]


The recurrence terminates when one of $u, v, w \le k$. It can be verified that 
$C(u,v,w) = O^*((u+v+w)^3)$. It then follows that $C(u,v,w) = O^*((u+v+w)^3)$ for any values of $u$, $v$, $w$, and this completes the proof of Theorem~\ref{thm:env4d}.
$\Box$


\section{Vertical Decomposition of Arrangements in $\reals^3$}
\label{sec:output}


Let $\S$ be a set of $n$ surfaces or surface patches in $\reals^3$ in general position, 
each of which is semi-algebraic of constant complexity, and let $X$ denote the 
number of vertices of $\A(\S)$.
For simplicity, and with no loss of generality, we assume that the 
surfaces are graphs of possibly partially defined continuous functions. This can be ensured
by cutting surfaces into surface patches at their silhouettes and at their curves of singularity.
We show that the complexity of the vertical decomposition of $\A(\S)$ is $O^*(n^2+X)$.

As in Section~\ref{sec:union}, it suffices to bound the number of
vertical visibilities between pairs of edges of $\A(\S)$. Again, we randomly color each surface as 
either red or blue, and only consider visibilities between red-red edges and blue-blue edges, in 
which the red-red edge lies below the blue-blue edge. An original vertical visibility has $1/8$
probability to appear as a visibility of the desired kind under the coloring scheme.
That is, up to a constant factor, the bound that we seek is also an upper bound for the original uncolored case.
Here too, each monochromatic edge $e$ may in general be the union of several original edges of 
$\A(\S)$. Therefore the number of these monochromatic edges is at most $O(X)$. As before, we denote 
the subsets of red surfaces and blue surfaces as $\R$ and $\B$, respectively, and put $m := |\R|$, 
$n := |\B|$, slightly abusing the notation, as above.

The high-level analysis proceeds more or less as in Section~\ref{sec:union}. That is, we apply a 
two-step partitioning scheme, in which we first partition the blue surfaces into $k$ subsets
$\B_1,\ldots,\B_k$, each of $n/k$ surfaces (in fact, the number of these surfaces in each subcell is 
at most $2n/k$---see below for the details of the analysis). Then, for each red-red edge $e$, we form $k$ 
separate lower envelopes of the blue surfaces, one for each $\B_i$, within the (upward) vertical curtain
erected from $e$, and analyze the complexity of the lower envelope of all these envelopes.

Denote by $C(m,n,X_1,X_2)$ the maximum number of vertical visibilities between red-red edges 
and blue-blue edges in an arrangement of a set $\R$ of at most $m$ red surfaces and a set $\B$
of at most $n$ blue surfaces, so that the complexity (number of vertices) of $\A(\R)$ is at most $X_1$ and the 
complexity of $\A(\B)$ is at most $X_2$. Observe that $X_1+X_2\le X$.

A major new aspect of the analysis is in handling the parameter $X$, now replaced by $X_1$ 
and $X_2$. The issue is that we have no control on how $X_1$ and $X_2$ are distributed over
the subproblems that arise when we partition $\B$ into $k$ arbitrary subsets, and then do the
same for $\R$, as we did in Section~\ref{sec:union}.

We overcome this issue by partitioning each of $\R$, $\B$ into $k$ \emph{random} subsets, say 
by choosing the subset to which a surface belongs independently and uniformly at random. Specifically, consider
the first partitioning step, where $\B$ is split. We form a random partition of $\B$ into $k$ 
subsets $\B_1, \ldots, \B_k$, where a surface $\sigma \in \B$ is assigned to a subset $\B_i$, 
$1\le i \le k$, which is chosen with probability $1/k$, independent of the assignment of the 
remaining surfaces in $\B$. This probabilistic model obeys the multinomial distribution with 
$k$ ``categories''. In particular, this implies that the size of each $\B_i$ is a binomial random variable 
with parameters $n$ and $1/k$. Similarly, when we apply such a random partition to $\R$ at the
second partitioning step, we obtain a partition into $k$ subsets $\R_1, \ldots, \R_k$, where the size of
each $\R_j$ is a binomially distributed random variable with parameters $m$ and $1/k$. We clearly 
have $E[|\B_i|] = n/k$, for each $1\le i \le k$, and $E[|\R_j|] = m/k$, for each $1\le j \le k$.

Using standard probabilistic arguments, exploiting the multiplicative Chernoff bound~\cite{AS-08}, 
we conclude that, with high probability, 
\begin{align*}
|\B_i| & \le n/k\left(1 + O\left(\sqrt{\frac{k}{n}\log{n}}\right)\right) ,\qquad\text{for each $1\le i \le k$, and} \\
|\R_j| & \le m/k\left(1 + O\left(\sqrt{\frac{k}{m}\log{m}}\right)\right) ,\qquad\text{for each 
$1\le j \le k$} .
\end{align*}
By choosing $k$ appropriately, we can assume that, with high probability,
these upper bounds do not exceed $2n/k$, and $2m/k$, respectively.


Moreover, at the first partitioning step, a blue-blue edge $e'$ is assigned to a specific subset
$\B_i$ with probability at most $1/k^2$ (here too, a blue-blue edge of $\A(\B_i)$ may be the union of several original edges of $\A(\B)$). 
Specifically, $e'$ is defined by at most four surfaces. That is, if $e'$ contains two endpoints (each of which is a vertex of the arrangement obtained by the intersection of a triple of surfaces) then this number is
four, if it has only one endpoint then $e'$ is defined by three surfaces, otherwise, it is defined by a pair of surfaces (recall that we exclude silhouette and singularity edges, in which case there is only a single surface defining an edge).

In the first two scenarios $\B_i$ has to contain the triple of surfaces defining an endpoint of $e'$ 
(or the quadruple defining both endpoints), which occurs with probability at most $1/k^3$. 
In the latter scenario the pair of surfaces defining $e'$ has to be assigned to $\B_i$, which happens with probability $1/k^2$.
Therefore the expected complexity of the arrangement $\A(\B_i)$ is $O(n^2/k^2 + X_2/k^3)$.
We comment that the events that edges show up in a specific subset are not independent. However, we claim below that, with high probability, the complexity of $\A(\B_i)$ is $O(n^2/k^2 + X_2/k^2)$, for each $1 \le i \le k$. This bound is slightly worse than the expected complexity, but it suffices for the analysis to proceed.

Indeed, since we have, with high probability, $|\B_i| \le 2n/k$, for each $1 \le i \le k$, we immediately conclude that the number of edges of $\A(\B_i)$ that are formed by pairs of surfaces is $O(n^2/k^2)$ (with high probability).
Regarding the number of edges that are formed by a triple (or a quadruple) of surfaces, their expected number $Y$ is $O(X_2/k^3)$, as observed above. Using Markov's inequality we conclude that the probability that the actual number of such edges exceeds $2kY$ is at most $1/(2k)$. That is, with probability at least $1 - 1/(2k)$, the number of such edges in $\A(\B_i)$ is at most $O(X_2/k^2)$. Using the probability union bound, we obtain that this bound holds for all sets $B_i$, $1 \le i \le k$, with probability at least $1/2$.
We comment that this event is conditioned on the event that $|\B_i| \le 2n/k$, for each $1 \le i \le k$ (which occurs with very high probability), so using the rule of conditional probability, we can assume that with probability at least $1/4$ the overall complexity of $\A(\B_i)$ is at most $O(n^2/k^2 + X_2/k^2)$, for each $1 \le i \le k$. By the probabilistic method~\cite{AS-08} this implies that there exists such a partition $\B_1, \ldots, \B_k$.

Hence, a suitable adaptation of the analysis in Section~\ref{sec:union}
yields the first-level recurrence (where $c > 0$ below is an absolute constant):
\[
C(m,n,X_1,X_2) \le k\beta(k) C(m, 2n/k, X_1, c(n^2/k^2 + X_2/k^2)) + k\beta(k) X_1 + O^*(mn) ,
\]
for a suitable near-constant extremely slowly growing function $\beta(k)$.
The overhead term $O^*(mn)$ comes from vertical visibilities that involve silhouettes and singularities, and follows by an argument similar to that in Section~\ref{sec:union}.

We next switch the roles of red and blue, and apply the same analysis to each pair $\R$, $\B_i$ of 
surfaces, keeping $\B_i$ fixed and partitioning $\R$ into $k$ random subsets, as above, each of which is of 
size at most $2m/k$ (with high probability). The analysis proceeds in a similar manner, and yields the bound 
\[
k^2\beta^2(k) C(2m/k, 2n/k, c(m^2/k^2 + X_1/k^2), c(n^2/k^2 + X_2/k^2)) + O_k(X_1 + X_2) + O^*_k(mn)
\]
on the number of vertical visibilities, where the $O_k(\cdot)$ notation indicates that the constant
of proportionality depends on $k$. That is, we obtain the recurrence
\[
C(m,n,X_1,X_2) \le k^2\beta^2(k) C(2m/k, 2n/k,  c(m^2/k^2 + X_1/k^2), c(n^2/k^2 + X_2/k^2)) + O_k(X_1 + X_2) + O^*_k(mn) 
\]
By choosing $k$ to be a sufficiently large constant, the solution of the recurrence is easily
seen to be 
\[
C(m,n,X_1,X_2) = O^*(m^2 + n^2 + X_1 + X_2) .
\]
That is, replacing $m$ and $n$ by the original value of $n$, and $X_1$, $X_2$ by the original
quantity $X$, we obtain the following:
\begin{theorem}
  \label{thm:arrg}
  Let $\S$ be a collection of $n$ constant-complexity semi-algebraic surfaces or surface patches
  in $\reals^3$, and let $X$ be the number of vertices in $\A(\S)$. Then 
  the complexity of the vertical decomposition of $\A(\S)$ is $O^*(n^2 + X)$.
\end{theorem}


\section{Constructing Cuttings and Decompositions}
\label{sec:alg}


\subsection{Constructing cuttings}
\label{sec:cutting}

Let $\S$ be a collection of $n$ semi-algebraic sets of constant complexity in $\reals^d$.
Let $\Pi$ be a substructure of $\A(\S)$, say, defined by a collection of cells of $\A(\S)$ that 
satisfy certain properties (e.g., lying in the complement of the union or lying below the lower envelope).
For a parameter $r>1$, a \emph{$(1/r)$-cutting} of $\Pi$
is a set $\Xi$ of pseudo-prisms with pairwise-disjoint relative interiors that cover $\Pi$, such that the
relative interior of each pseudo-prism $\tau \in \Xi$ is crossed by (intersected by but not contained in) at 
most $n/r$ sets of $\S$. The subset of $\S$ crossed by $\tau$ is called the \emph{conflict list} of $\tau$.

It is well known that the random-sampling paradigm can be used to construct 
a $(1/r)$-cutting~\cite{Ag:rs,dBS-95,HW87,m-ept-92}. Namely, set $s=cr\log r$, where $c$ is a sufficiently 
large constant. Let $\R \subseteq \S$ be a random subset of $\S$ of size $s$, and let $\VD(\R)$ be the vertical decomposition of $\A(\R)$. For each cell $\tau\in\VD(\R)$, let 
$\S_\tau \subset \S$ be the subset of $\S$ that crosses $\tau$. By construction, $\S_\tau \cap \R = \emptyset$
and $\tau$ is a semi-algebraic set of 
constant complexity, therefore using a standard random-sampling argument~\cite{Clarkson-87,HW87}, it can be shown 
that $|\S_\tau| \le n/r$ for all $\tau\in\VD(\R)$ with probability at least $1/2$ assuming the constant $c$ is chosen sufficiently large. Therefore, to construct a $(1/r)$-cutting $\Xi$ of $\Pi$, 
we only have to decide which of the cells of $\VD(\R)$ should be 
included in $\Xi$ to ensure that they cover $\Pi$.

If $\S$ is a set of semi-algebraic sets in $\reals^3$ and we wish to compute a $(1/r)$-cutting of $\C(\S)$, 
the complement of the union of $\S$, we set $\Xi = \{ \tau \in \VD(\R) \mid \tau \subseteq \C(\R)\}$. 
Since $\R \subseteq \S$, $\C(\S) \subseteq \C(\R)$, and thus $\Xi$ is guaranteed to cover $\C(\S)$. By 
Theorem~\ref{thm:main}, $|\Xi| = O^*(r^2+U(r))$. In contrast, if we want to construct a $(1/r)$-cutting of the entire $\A(\S)$, we set $\Xi=\VD(\R)$. If $\A(\S)$ has $X$ vertices, then the expected number of vertices 
in $\A(\R)$ is $O(r^2+Xr^3/n^3)$, and thus, by Theorem~\ref{thm:arrg},
the expected size of $\Xi$ is $O^*(r^2+Xr^3/n^3)$. (If the size of $\Xi$ is more than twice its expected size, we discard $\Xi$ and repeat the construction.) 
Finally, if $\S$ represents graphs of a set of trivariate functions in $\reals^4$
and we wish to construct a $(1/r)$-cutting of the portion of $\A(\S)$ lying below
the lower envelope of $\S$, we set $\Xi$ to be the set of cells of 
$\VD(\R)$ that lie below the lower envelope of $\R$. By Theorem~\ref{thm:env4d}, $|\Xi| = O^*(r^3)$. 
Hence, we conclude the following:\footnote{%
It is possible to reduce the size of the cuttings by a 
polylogarithmic factor using a two-level sampling scheme as described 
in~\cite{AMS98,dBS-95,cf-dvrsi-90,m-ept-92}. Since we are 
using $O^*()$ notation and are ignoring subpolynomial factors, we described a simpler, albeit 
slightly weaker, construction.}

\begin{theorem}
  \label{thm:cuttings}
	\begin{itemize}
		\item[(i)] Let $\S$ be a collection of $n$ semi-algebraic sets of constant complexity in $\reals^3$, and let 
  $U(m)$ be an upper bound on the complexity of the union of at most $m$ objects of $\S$.
  There exists a $(1/r)$-cutting of $\C(\S)$, the complement of the union of $\S$, of size $O^*(r^2 + U(r))$.
  
  \item[(ii)] Let $\F$ be a collection of $n$ trivariate semi-algebraic functions of constant complexity. 
  There exists a $(1/r)$-cutting of the region below the lower envelope of $\F$ of size $O^*(r^3)$.
  
  \item[(iii)] Let $\S$ be a collection of $n$ constant-complexity semi-algebraic surfaces or surface patches
  in $\reals^3$, so that the number of vertices in $\A(\S)$
  is $X$.
  Then there exists a $(1/r)$-cutting for $\S$ of size $O^*(r^2 +  r^3X/n^3)$.
	\end{itemize}
For contestant values of $r$, these cuttings, along with the conflict lists of their cells,
can be computed in $O(n)$ expected time (where the constant of proportionality depends on $r$).
\end{theorem}


\subsection{Constructing vertical decompositions}
\label{subsec:alg}


We now describe algorithms for constructing vertical decompositions for the cases studied in Sections~\ref{sec:union}--\ref{sec:output}.


\paragraph{Complements of unions in $\reals^3$.}
Let $\S$ be a collection of $n$ semi-algebraic sets (each of constant complexity) in $\reals^3$ such that 
the maximum complexity of the union of any subset $\S'$ of $\S$ of at most $m \le n$ sets is $U(m)$. 
Let $\C(\S')$ denote the complement of $\U(\S')$.

We present below an algorithm that constructs, in $O^*(n^2+U(n))$ expected time,
the vertical decomposition of $\C(\S)$.
More precisely,  it constructs the set of pseudo-prisms in the  
vertical decomposition of $\C(\S)$.
As a main step of the algorithm, we perform the subtask of reporting 
all the vertical visibilities, within $\C(\S)$, between pairs of edges $(e, e')$ that lie on $\bd{\U(\S)}$. 
By Theorem~\ref{thm:main}, the number of these vertical visibilities is $O^*(n^2+U(n))$. 
We also compute the vertices, edges, and $2$-faces of $\U(\S)$ in $O^*(n^2+U(n))$ expected time, e.g., using the randomized incremental algorithm described in~\cite{AAS-97}. Then the pseudo-prisms in 
$\VD(\S)$ can be computed in a fairly standard (though somewhat tedious) manner by traversing all the 
faces and edges of $\bd{\U(\S)}$ and tracking their vertical visibilities.
We omit the details from here in the interest of brevity, and refer the reader to~\cite{dBGH}, where
a similar method was used for computing the vertical decomposition of an arrangement of triangles in $\reals^3$.

We follow a randomized divide-and-conquer scheme to compute vertical visibilities.
Let $1 \le r \le n$ be a sufficiently large constant parameter. 
If $|\S| \le n_0$, where $n_0$ is a constant that depends on $r$, we 
report all pairs of vertical visibilities between the edges on $\bd{\C(\S)}$ in a brute-force manner. 
Otherwise, we recursively construct a $(1/(2r))$-cutting $\Xi$ of $\C(\S)$ of size $O^*(r^2+U(r))$, using 
Theorem~\ref{thm:cuttings}~(i). (We comment that the actual reporting is done only at the bottom of the recurrence.)
%
For each cell $\tau\in\Xi$, let $\S_\tau \subset \S$ be its conflict list, the family of input sets that cross the relative interior of $\tau$, plus the $O(1)$ input sets that define the cell $\tau$.  
By construction, $|\S_\tau| \le n/(2r)+O(1) \le n/r$.
As is easily verified, any edge pair $(e,e')$ (that lie on $\bd{\C(\S)}$) of vertical visibility 
within $\C(\S)$ must be reported during this process, since the vertical segment $\rho$ connecting $e$ and 
$e'$ must be contained in some prism cell of $\Xi$. Otherwise, this would imply that 
one of the input sets crosses $\rho$, but this violates the definition of vertical visibility. 
The overall expected running time $T(n)$ to report all pairs of vertical visibility obeys the recurrence:
$$
T(n) = O^*(r^2+U(r)) T(n/r) + O^*(n) , 
$$
where the overhead term accounts for computing $\Xi$ and the conflict lists of all the cells of $\Xi$.
Using induction, it can be verified that the solution is $T(n) = O^*(n^2+U(n))$.
We have thus shown:
\begin{theorem}
  \label{thm:algo}
  Let $\S$ be a collection of $n$ constant-complexity semi-algebraic sets in $\reals^3$, such that the
	complexity of the union of any subset of $\S$ of size $m$ is $U(m)$.
	Then the vertical decomposition of $\C(\S)$ can be constructed in $O^*(n^2+U(n))$ 
	randomized expected time.
\end{theorem}


\paragraph{Arrangements in $\reals^3$.}

Let $\S$ be a collection of $n$ semi-algebraic sets (each of constant complexity) in $\reals^3$ such that 
$\A(\S)$ has $X$ vertices.
The above approach for computing the vertical decomposition of $\C(\S)$
can be extended to compute the vertical decomposition of $\A(\S)$. The only difference is 
that we now compute a $(1/(2r))$-cutting of $\A(\S)$ of size $O^*(r^2+r^3X/n^3)$ using Theorem~\ref{thm:cuttings}~(iii). 
Omitting the straightforward details, we conclude the following result.
\begin{theorem}
  \label{thm:arr-algo}
  Let $\S$ be a collection of $n$ constant-complexity semi-algebraic sets in $\reals^3$ such that
  the arrangement $\A(\S)$ has $X$ vertices.
  Then the vertical decomposition of $\A(\S)$ can be constructed in $O^*(n^2+X)$ 
  randomized expected time.
\end{theorem}


\paragraph{Lower envelopes in four dimensions.}
Let $\F$ be a collection of $n$ trivariate semi-algebraic 
functions of constant complexity. Our goal is to construct the vertical decomposition
of $E^-$, the portion of $\A(\F)$ lying below the lower envelope $E$ of $\F$.


We briefly recall how the vertical decomposition is defined. We iterate over the functions of $\F$.
For each function $a\in\F$, we form the 2D intersection surfaces $a\cap b$, for $b\in\F\setminus\{a\}$, which we
denote for short as $ab$. We project these surfaces onto the $xyz$-space, and construct the vertical decomposition
of the complement $\C_a$ of the union $\U_a$ as defined in Section~\ref{sec:env}. As in the basic construction in
Section~\ref{sec:union}, the key step is to find all the vertical visibilities within $\C_a$. Each such 
visibility is between two edges, each of which is the intersection of two of the surfaces $ab$ (for $a$ fixed).
We denote for short the intersection curve of $ab$ and $ac$ as $abc$. That is, we need to find all the 5-tuples 
$(a,b,c,d,e)$ of distinct functions of $\F$, such that $abc$ and $ade$ form a vertical visibility (in the 
$z$-direction) within $\C_a$. Once we have found all these 5-tuples, completing the representation of the
vertical decomposition can be carried out in a routine manner, similar to that used in the three-dimensional
case reviewed earlier, which, for this setting, takes overall $O^*(n^3)$ time.

To construct the above visibilities, we proceed as above. Namely, we construct a $(1/(2r))$-cutting $\Xi$ 
of $E^-$ of size $O^*(r^3)$ using Theorem~\ref{thm:cuttings}~(ii).
For each prism $\tau\in \Xi$, let $\F_\tau$ be its conflict list plus the $O(1)$ functions that define $\tau$.
We process recursively each prism cell $\tau$, where at the bottom of the recursion we report all 
pairs of vertical visibilities between the edges of $\F_\tau$ in a brute force manner. 

We claim that, for each vertical visibility (in the full collection $\F$) defined by a 5-tuple $(a,b,c,d,e)$,
all five functions appear in the conflict list of the same prism $\tau\in \Xi$, 
so the visibility will be found in the corresponding recursive step (in fact, as just described, it will be found at some leaf of the recursion).
Indeed, let $\zeta$ be the $z$-vertical segment in the $xyz$-space that defines the visibility, with endpoints
on $abc$ and on $ade$. Let $\zeta^+$ be the lifting of $\zeta$ to the graph of $a$. Then $\zeta^+$ is 
fully contained in $E_a$, and in fact no function graph crosses the downward vertical curtain erected 
(in the $w$-direction) from $\zeta^+$. 

We claim that $\zeta^+$ is fully contained in a prism $\tau\in \Xi$, 
from which the previous claim follows readily.
Suppose to the contrary that this is not the case, so $\zeta^+$ crosses the boundary of such a prism. 
Since $\zeta^+$, or rather $\zeta$, is in the $z$-direction, it follows that $\zeta$ must hit the 
floor or the ceiling, in the $z$-direction, of a prism of the three-dimensional decomposition of the minimization
diagram, which, by construction, lies on some ($xyz$-projection of an) intersection surface, say $uv$. This however
is impossible, since no such surface can cross the interior of $\zeta$, which is fully contained in $\C_a$, which is
disjoint from all such surface projections. 
%
This establishes the correctness of the procedure and yields the following:

\begin{theorem}
  \label{thm:env_algo}
  Let $\F$ be a collection of $n$ trivariate semi-algebraic functions of constant complexity.
  Then the vertical decomposition of portion of $\A(\F)$ lying below
  the lower envelope of $\F$ 
  can be constructed in randomized expected time $O^*(n^3)$.
\end{theorem}

\section{Output-Sensitive Point-Enclosure Reporting in $\reals^3$}
\label{sec:point_enclosure_cyl}

In the problem considered in this section we have a set $\S$ of $n$ semi-algebraic regions of constant complexity
in $\reals^3$, with a bound $U(m)$ on the complexity of the union of any subset of at most $m$ regions of $\S$.
As in the earlier sections we assume here that $U(m) = O^*(m^2)$.

The goal is to preprocess $\S$ into a data structure that can support \emph{output-sensitive point enclosure 
reporting} queries, where a query specifies a point $p$ and seeks to report all the regions of $\S$ that contain $p$.
We present an algorithm that uses $O^*(n^2)$ preprocessing and storage, and answers a query is $O^*(1+k)$ time,
where $k$ is the output size.

The technique that we present bears some resemblance to the original technique of Matou\v{s}ek~\cite{Ma:rph},
which has been developed for the case where $\S$ consists of halfspaces, but is different in several key aspects.


Let $k$ be a parameter. We take a random sample $\R$ of $\frac{\alpha n}{k}$ regions from 
$\S$ (in expectation), by choosing each region independently with probability $q = \frac{\alpha}{k}$, for some 
small constant $0<\alpha<1$, and construct the vertical decomposition $\VD(\R)$ of the complement 
$\K = \K(\R)$ of the union of $\R$. As shown in Section~\ref{subsec:alg}, 
this takes time $O^*((n/k)^2)$. In addition, we associate with
each cell $\tau$ of $\VD(\R)$ its \emph{conflict list} $\S_\tau$, which is the set of those regions 
of $\S$ whose boundary crosses $\tau$, and the set $\S_\tau^0$ of those regions of $\S$ that fully 
contain $\tau$. By random sampling theory, with high probability, 
\[
|\S_\tau| = O(k\log(n/k)) = O(k\log n)
\]
for every cell $\tau$.

To obtain an efficient construction of the conflict lists, as well as an efficient procedure for
locating the points of $P$ in $\VD(\R)$ (see below), we modify the construction of $\VD(\R)$ and 
make it \emph{hierarchical}, as follows. We construct a hierarchical tree structure of decompositions.
At each recursive step we take a random sample $\R_0$ of $r_0$ regions from the current set
$\S'$, for a sufficiently large constant parameter $r_0$, and construct the vertical decomposition
$\VD(\R_0)$ of the complement of the union of $\R_0$. For each cell $\tau$ of $\VD(\R_0)$ we construct
its conflict list $\S'_\tau$, and the set $\S^{'0}_\tau$ of regions that fully contain $\tau$.
Since $r_0$ is constant, this takes $O(|\S'|)$ time over all cells $\tau$. We recursively repeat 
the construction for each conflict list $\S'_\tau$. The construction terminates when we reach 
subproblems with $O(k\log n)$ regions.

A single recursive step in this hierarchy with $n$ input regions generates $O^*(r_0^2)$ subproblems, each with
$O^*(n/r_0)$ regions. The overhead nonrecursive cost of the step is $O(n)$, where the constant of 
proportionality depends on $r_0$. Hence, at level $j$ of the recursion, we have $O^*(r_0^{2j})$ subproblems,
each of size $O^*(n/r_0^j)$, and the total cost of constructing the whole structure,
taking also into account the overhead costs, is therefore
\[
O^*\left( \sum_j r_0^{2j}\cdot \frac{n}{r_0^j} \right) = 
O^*\left( n \sum_j r_0^j \right) = O^*\left( n r_0^{j_{\rm max}} \right) , 
\]
where the constant of proportionality is the product of two factors, one depending on $r_0$ but not
on $j$, and one of the form $c^j$ for some absolute constant, independent of $r$, and where $j_{\rm max}$
is the maximum level of recursion, which satisfies $n / r_0^{j_{\rm max}} \approx k\log n$, or 
$r_0^{j_{\rm max}} = O^*(n/k)$. That is, the overall cost of the construction is $O^*(n^2/k)$.

Note that the output of the hierarchical construction is not necessarily $\VD(\R)$, but it 
suffices for our need. Specifically, we now locate the query point $p$ in the structure. 
At each node $\tau$ that the search reaches, we find, in brute force, 
the cell $\tau'$ of the local vertical decomposition at $\tau$ that contains $p$, and 
continue the search recursively at $\tau'$. At each step of the search, at any node $\tau$, 
we can report all the regions of $\S^0_\tau$, as they certainly contain $p$. If at some 
step we detect that $p$ does not lie in any cell of the local vertical decomposition, we 
conclude that $p$ does not belong to $\K$, and terminate the search (see below for the rationale of this termination). Otherwise the search reaches a leaf $\tau$. We then report all the regions in $\S^0_{\tau'}$, over all nodes
$\tau'$ along the search path. The only unreported regions that might contain $p$ are 
those in $\S_\tau$. As will follow from the overall structure of the algorithm, we can afford 
to inspect all the regions of $\S_\tau$, and output those among them that contain $p$. The cost 
of the query is $O(\log n)$ plus a cost that, as we will show, is larger than the number of regions 
that have been reported, by at most a logarithmic factor.

\paragraph{The full reporting procedure.}
Let us first consider the offline problem, where we are given $m$ query points.
Define the \emph{depth} of a point $p$ to be the number of regions of $\S$ that contain $p$. 
Let $p$ be a point of depth $j\le k$. Then the probability that $p$ belongs to $\K(\R)$ is 
\[
(1-q)^j \ge (1-q)^k \approx e^{-qk} = e^{-\alpha} .
\]
We can make this probability very close to $1$ by
independently drawing $c \ln{m}$
random samples $\R$. The probability that $p$ avoids the complement of the 
union for all these samples is at most
\[
\left( 1 - e^{-\alpha} \right)^{c\ln m} \approx e^{-ce^{-\alpha}\ln m} = \frac{1}{m^{ce^{-\alpha}}} .
\]
By choosing $c$ sufficiently large, we can ensure that, with high probability, all query points, 
taken from some set of $m$ possible queries, are captured in this manner (i.e., belong to $\K(\R)$
for at least one sample $\R$). 

We now construct a geometric sequence of these structures, for $k=2^j$ with $j=1,2,\ldots$.
For a point $p\in P$, let $j$ be the first index for which $p$ lies in a cell $\tau$ of one of 
the vertical decompositions constructed for $k = 2^j$. (Informally, as $j$ increases, the size 
of the sample $\R$ decreases, so $\K(\R)$ increases, making it `easier' for $p$ to belong to $\K(\R)$.
In the worst case, $p$ may stay inside the union for every $k$, but the analysis will handle
this case too.) The preceding analysis implies that, with high probability, the depth of $p$ is 
larger than $k/2$ (otherwise $p$ would have been captured earlier, with high probability).
But then we can afford to inspect all the $O(k\log n)$ regions in the conflict list of $\tau$ and report those that contain $p$, in the sense that the size of the list is larger than the output size by at most a logarithmic factor.

The cost of constructing all these structures is $\sum_j O^*(n^2/2^j) = O^*(n^2)$, and 
the cost of locating the query point in the respective vertical decompositions, ignoring the
reporting part of the cost, is $O^*(1)$. For the reporting part, we do not report anything when
we find out that the query point $p$ is not in the current vertical decomposition. 
At the first time when $p$ lies in $\VD(\R)$, for each cell $\tau'$ that it visits, all the 
regions of $\S^0_{\tau'}$ are reported. Then at the leaf $\tau$ that the search reaches, we
iterate over its conflict list and report those regions that contain $p$. The overall
reporting cost is proportional to the output size, up to a logarithmic factor, which may 
arise when we iterate over the conflict list of the leaf, which may be larger than the
output size by a logarithmic factor. That is, we have a data structure of size $O^*(n^2)$,
where each point-enclosure query costs $O^*(1+k)$. Hence, if we want to perform $m$ point-enclosure reporting 
queries, the overall cost, including preprocessing, 
is $O^*(n^2 + m + K)$, where $K$ is the overall output size.


Returning to the online problem (where queries are given online),
we observe that there are at most $O(n^3)$ combinatorially different queries, each of which corresponds to a cell in the arrangement $\A(\S)$. Therefore, we follow verbatim the above analysis with $m = O(n^3)$.
This results in a data structure of overall storage and preprocessing $O^*(n^2)$, for the case where
$U(n) = O^*(n^2)$, or $O^*(U(n))$ otherwise, which answers point-enclosure reporting queries in $O^*(1+k)$ time.

\paragraph{Point enclosure reporting in 4D.}
The same machinery can be used for performing output-sensitive point-enclosure reporting queries in $\reals^4$. In this setup
we have a collection $\F$ of $n$ semi-algebraic trivariate functions of constant complexity, and we want to 
preprocess $\F$ into a data structure, so that, for a query point $q\in\reals^4$, it reports all the functions 
of $\F$ whose graphs pass below $q$. A more or less identical analysis shows that this can be done 
with $O^*(n^3)$ storage and preprocessing time, so that a query of the above kind can be answered in 
$O^*(1+k)$ time, where $k$ is the output size, namely the number of functions below the query point. 
Hence, in an offline version, $m$ queries can be answered in $O^*(n^3+m+K)$ time, including preprocessing, 
where $K$ is the overall output size.

In conclusion, we have shown:
\begin{theorem}
  \label{thm:enc}
  (a) Let $\S$ be a set of $n$ semi-algebraic regions of constant complexity in $\reals^3$, so that 
  the complexity of the union of any subset of at most $m$ regions of $\S$ is $O^*(m^2)$.
  Then $\S$ can be preprocessed into a data structure of size $O^*(n^2)$, in $O^*(n^2)$ randomized 
  expected time, which supports point-enclosure reporting queries in time $O^*(1+k)$, where each
  query is with a point $q\in\reals^3$, and seeks to report all regions of $\S$ that contain $q$,
  and $k$ is the output size.

  \smallskip\noindent
  (b) Let $\F$ be a set of $n$ semi-algebraic trivariate functions of constant complexity. 
  Then $\F$ can be preprocessed into a data structure of size $O^*(n^3)$, in $O^*(n^3)$ randomized 
  expected time, which supports point-enclosure queries in time $O^*(1+k)$, where each query
  is with a point $q\in\reals^4$, and seeks to report all functions of $\F$ whose graphs pass
  below $q$, and $k$ is the output size.
\end{theorem}



\section{Nearest Neighbor Searching amid Lines in $\reals^3$}
\label{sec:plnn}


We now turn our attention to nearest-neighbor-searching problems involving points and lines in $\reals^3$. In 
this section, we present a linear-size data structure for preprocessing a 
set $L$ of $n$ lines in $\reals^3$ into a data structure so that 
for a query point $p \in \reals^3$, the line of $L$ nearest to $p$ can be reported quickly (more quickly
than what can be obtained by the standard machinery). 
Using standard techniques (e.g.,
parametric search)~\cite{Ag:rs,AM}, a nearest-neighbor query, referred to as an NN query on $L$, can be reduced to answering
$O^*(1)$ \emph{sphere-intersection-detection} queries on $L$. That is, we want to preprocess $L$ into a 
data structure that can efficiently determine whether a query sphere $\sigma$ intersects any of the lines in $L$.



\begin{figure}[htb]
   \begin{center}
     {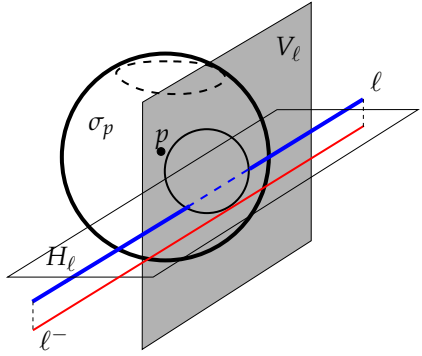 } 
  \end{center}
  \caption{Illustration of condition (ii-) of Lemma~\ref{lem:lower}. Here $\ell$ is lower than $\sigma_p$.}
  \label{fig:point_line_NN}
\end{figure}

\paragraph{Overall data structure.} Our overall data structure is based on the following technical property,
originally proved  by Mohaban and Sharir~\cite{MS}.
Let $\ell$ be a line in $\reals^3$, and let $\sigma_p$ be a sphere, centered at a point $p$.
Let $V_\ell$ be the vertical plane that contains $\ell$, and let $H_\ell$ be the plane
that contains $\ell$ and is orthogonal to $V_\ell$.
We say that $\ell$ is \emph{lower} (resp., \emph{higher})
than $\sigma_p$ if $p$ lies above (resp., below) $H_\ell$;
see Figure~\ref{fig:point_line_NN}.

\begin{lemma}[\cite{MS}]
  \label{lem:lower}
	Assuming that $\ell$ is lower than $\sigma_p$ (using the above notation), $\ell$ intersects $\sigma_p$
  if and only if the following two conditions hold:
  \begin{description}
  \item[(i)]
    The $xy$-projections of $\ell$ and of $\sigma_p$ intersect, and
  \item[(ii-)]
    $\ell$ lies above the parallel line $\ell^-$ that lies in $V_\ell$ and is tangent to $\sigma_p$ from below.
  \end{description}
  Symmetrically, assuming that $\ell$ is higher than $\sigma_p$, $\ell$ intersects $\sigma_p$
  if and only if (i) holds and
  \begin{description}
  \item[(ii+)]
    $\ell$ lies below the parallel line $\ell^+$ that lies in $V_\ell$ and is tangent to $\sigma_p$ from above.
  \end{description}
\end{lemma}

We describe a linear-size data structure that, for a query sphere $\sigma$, determines whether any line of $L$ 
that is lower than $\sigma$ intersects $\sigma$. (A similar data structure can be constructed for detecting 
whether any line of $L$ that is higher than $\sigma$ intersects $\sigma$.)  We thus need a data structure 
that, for a query sphere $\sigma$, returns \textsc{Yes} if a line in $\ell$ satisfies the following three 
conditions, as in Lemma~\ref{lem:lower}: 
\begin{itemize}
	\item[(C1)] the $xy$-projections of $\ell$ and $\sigma$ intersect, 
	\item[(C2)] $\ell$ is lower than $\sigma$, and 
	\item[(C3)] $\ell$ lies above the parallel line $\ell^-$ that lies in $V_\ell$ and is tangent to 
		$\sigma_p$ from below.
\end{itemize}


We use a multi-level partition tree~\cite{Ag:rs,AAEKS} for answering queries of this kind. 
In particular, we construct a $3$-level partition tree, each of whose nodes $v$ stores a ``canonical'' 
subset $L_v \subseteq L$. The first-level tree identifies the subset of lines that satisfy condition (C1)
for the given query. Since a line 
in $\reals^2$ requires two parameters, (C1) can be formulated as a two-dimensional semi-algebraic 
range query of a very simple nature---the inequality that we need to test just involves the absolute value 
of a linear expression. Thus the first level is a $2$-dimensional partition tree for semi-algebraic range 
queries of this simple kind~\cite{AMS,MP}. As shown in~\cite{MS}, and easy to see, (C2) just amounts to
testing whether the center of the sphere lies above the respective planes $H_\ell$, so it can be formulated 
as a $3$-dimensional halfspace range query. 
For each node $u$ of the first-level tree, we construct a $3$-dimensional partition tree for halfspace range
searching, on the subset of 
lines $L_u$ associated with $u$, as a second-level tree. Finally, for each node $v$ of every second-level 
tree, we construct a third-level partition tree on $L_v$, the subset of lines associated with $v$,
which tests for (C3). We present below a linear-size data structure that can test condition (C3) in $O^*(n^{2/3})$ time (actually, in $O^*(|L_v|^{2/3})$ time). 
For a query sphere $\sigma$, the first two levels of the partition tree return the subset of lines 
that satisfy conditions (C1) and (C2) as the union of a few canonical subsets (see below for a
precise statement). For each of these canonical subsets $L_v$, the third-level tree 
constructed on $L_v$ is used to test whether any line in $L_v$ satisfies (C3). 
If the answer is \textsc{Yes}, then we conclude that $\sigma$ intersects a line of $L_v$ and return \textsc{Yes}.
Since the query time at each level is $O^*(n^{2/3})$ (it is actually smaller for the first level), 
the properties of multi-level partition trees 
(see, e.g., Theorem~A.1 in the appendix of \cite{AAEKS}), imply that the overall query time 
is also $O^*(n^{2/3})$. The overall size of the data structure is $O(n)$.\footnote{%
  A straightforward 
  application of the multi-level data-structure framework leads to a data structure of size $O^*(n)$.
  But, using well known machinery,
  the size can be improved to $O(n)$ while keeping the query time $O^*(n^{2/3})$ by constructing 
  secondary structures only at some of the nodes.
}

\paragraph{Sphere-intersection query for lines lower than the sphere.}

Let $L$ be a set of $n$ lines in $\reals^3$. We wish to preprocess $L$ into a linear-size data structure 
that, for a query sphere $\sigma$ satisfying conditions (C1) and (C2) for all lines in $L$, can 
determine in $O^*(n^{2/3})$ time whether $\sigma$ intersects any line of $L$.
We work in the $4$-dimensional parametric space of lines, denoted by $\Lspace$, where a line $\ell$ is 
represented by the point $\ell^* = (a,b,c,d)$ and the equations defining $\ell$ are $y=ax+c$, $z=bx+d$; 
$\Lspace$ is thus identified\footnote{%
  For convenience (and with no loss of generality if one assumes general position), 
  we ignore the fact that this space is actually projective.
}
with $\reals^4$. Put $L^* = \{ \ell^* \mid \ell \in L\}$. 
A sphere $\sigma$ is associated with a surface (patch) $\gamma_\sigma \subset \Lspace$, 
which is the locus of points $\ell^*$ such that the corresponding line $\ell$ is tangent to $\sigma$ from below. 
Let $\gamma_\sigma^+$ be the set of points lying on or above $\gamma_\sigma$ in the $d$-direction; 
$\gamma_\sigma^+$ is a semi-algebraic set of constant complexity.
It is easily seen that a line $\ell$ satisfying conditions (C1) and (C2) intersects $\sigma$ 
if and only if $\ell^*$ lies in $\gamma_\sigma^+$. Let $\Gamma$ be the collection of all sets 
$\gamma_\sigma^+$ such that $\sigma$ satisfies (C1) and (C2) for all lines in $L$.
Thus the sphere-intersection query for a sphere $\sigma$ in our setting reduces to 
semi-algebraic range-emptiness  query in $L^*$ with $\gamma_\sigma^+\in\Gamma$.
Using the known and standard partition tree mechanism~\cite{AMS,MP}, this query can be 
answered in $O^*(n^{3/4})$ time, but we show how to improve the query time to $O^*(n^{2/3})$.

We follow the approach of Matou\v{s}ek~\cite{Ma:rph} and of Sharir and Shaul~\cite{ShSh} 
for answering the range-emptiness query. We need a couple of definitions. 
Let $P \subset \Lspace$ be a set of $n$ points.  For a parameter $k\ge 0$, we call a semi-algebraic set 
$\gamma \subset \Lspace$, which semi-unbounded in the negative $d$-direction,
\emph{$k$-shallow} if $|P \cap \gamma| \le k$. For a parameter $r\ge 1$, we 
call a family $\Pi = \{(P_1, \Delta_1),\ldots, (P_u, \Delta_u)\}$ a $(1/r)$-partition for $P$
if (i) $\{P_1, \ldots, P_u\}$ is a partition of $P$, (ii) $n/2r \le |P_i| \le n/r$, and (iii)
$P_i \subset \Delta_i$ where $\Delta_i \subseteq \Lspace$ is a semi-algebraic set of constant 
complexity, referred to as a cell of $\Pi$. 
The \emph{crossing number} of $\Pi$ for a semi-algebraic set $\tau$, denoted by $\chi(\Pi,\tau)$,
is the number of cells of $\Pi$ intersected by the boundary of $\tau$. 
The crossing number of $\Pi$ for a family $\Xi$ of semi-algebraic sets, denoted by $\chi(\Pi,\Xi)$, is 
defined as $\max_{\tau\in\Xi} \chi(\Pi,\tau)$.

A major ingredient of the approach in~\cite{Ma:rph,ShSh} is to construct a 
so-called \emph{test set} $\Phi$ of a small number 
of semi-algebraic sets, which represent well all query semi-algebraic sets that are shallow. The following lemma of Sharir and Shaul~\cite[Theorem 3.2]{ShSh} summarizes the key property:
\begin{lemma}[\cite{ShSh}] 
	\label{lem:ms}
Let $P$ be a set of $n$ points in $\reals^d$, for some $d\ge 1$, and 
	let $\Gamma$ be a (possibly infinite) family of 
semi-algebraic sets of constant complexity.
Let $r\ge 1$ be a parameter, and let $\Phi$ be another finite collection (not necessarily a subset of 
$\Gamma$) of semi-algebraic sets of constant complexity with the following properties: 
\begin{description}
\item[(i)]
  Every set in $\Phi$ is $(n/r)$-shallow with respect to $P$. 
\item[(ii)]
  The complement of the union of any $m$ sets of $\Phi$ can be decomposed 
  into at most $\zeta(m)$ ``elementary cells'' (semi-algebraic sets of constant complexity) for any $m\ge 1$, 
  where $\zeta(m)$ is a suitable monotone increasing superlinear function of $m$. 
\item[(iii)]
  Any $(n/r)$-shallow set $\gamma\in \Gamma$ can be covered by the union 
  of at most $\delta$ ranges of $\Phi$, where $\delta$ is a constant (independent of $r$).
\end{description}
Then there exists a $(1/r)$-partition $\Pi$ of $P$ such that for any $(n/r)$-shallow range $\gamma \in \Gamma$, $\chi(\Pi,\gamma) = O(r/\zeta^{-1}(r) + \log r \log |\Phi|)$ if 
$\zeta(r)/r^{1+\eps}$  is monotonically increasing for some (arbitrarily small) 
constant $\eps > 0$, and $\chi(\Pi,\gamma) = O(r \log r/\zeta^{-1}(r) + \log r \log |\Phi|)$ otherwise.
Furthermore, $\Pi$ can be constructed in $(|\Phi| + n) r^{O(d)}$ expected time assuming $\Phi$ is given.
\end{lemma}

As shown in~\cite{Ma:rph,ShSh}, using Lemma~\ref{lem:ms} and assuming that $|\Phi|=r^{O(d)}$, one can 
construct a partition tree of linear-size that can determine in $O^*(n/\zeta^{-1}(n))$ time whether 
$\gamma\cap P \ne \emptyset$, for any query range $\gamma \in \Gamma$. 
We present an algorithm below for constructing a test set $\Phi$ of size $r^{O(1)}$ for our setup 
so that $\zeta(m) = O^*(m^3)$ and $\delta=1$, which in turn yields a linear-size data structure for 
sphere intersection queries with  $O^*(n^{2/3})$ query time, as desired.

\paragraph{Constructing a test set.}
To construct the test set, we also use the $4$-dimensional parametric space $\Sspace$ 
of spheres in $\reals^3$, where a sphere $\sigma$ of radius $r$  centered at a point $p$ 
is mapped to the point $\sigma^* = (p,r)\in\Sspace$; $\Sspace$ can thus be identified with $\reals^4$.
A line $\ell$ in $\reals^3$ is mapped to a surface
$\omega_\ell$, consisting of all points $\sigma^* \in \Sspace$ that represent spheres that touch 
$\ell$ from above. As is easily verified, these surfaces are monotone over the $xyz$-subspace, so that a point
$\sigma^*$ lies above the surface $\omega_\ell$ if and only if $\ell$ intersects $\sigma$,
assuming $\sigma$ and $\ell$ satisfy (C1) and (C2).\footnote{%
  Informally, this is why we have to distinguish between lines that pass below the sphere and lines that pass above.}

Let $\Omega = \{ \omega_\ell \mid \ell \in L\}$ denote the collection of these surfaces. We take a 
random subset $R\subseteq \Omega$ of $s = cr\log r$ surfaces, for some sufficiently large constant $r$, 
and construct the vertical decomposition $\VD(R)$ of the arrangement $\A(R)$;
$\VD(R)$ has $O^*(r^4)$ cells~\cite{Koltun-04a}. By a standard random-sampling argument~\cite{HW87},
each cell of $\VD(R)$ is crossed by at most $n/r$ surfaces of $\Omega$ with probability at least $1/2$. 
If this is not the case, we discard $R$ and choose another random subset, until we find one with
the desired property. We choose a subset $\Xi$ of $\VD(R)$, namely, those cells that have at most 
$n/r$ surfaces of $\Omega$ passing \emph{fully below} them. By construction, these cells cover the 
lowest $n/r$ levels of $\A(\Omega)$, and are contained in the at most $2n/r$ lower levels of $\A(\Omega)$.

Let $\tau$ be a cell of $\Xi$.
We now switch to the parametric line-space $\Lspace$, where each point $\sigma^*\in\tau$
becomes the surface $\gamma_\sigma$. We construct the lower envelope of the (infinitely many)
surfaces $\gamma_\sigma$ over all $\sigma^*\in\tau$. Let $\Phi_\tau\subset \Lspace$ be the set of points 
lying above the lower envelope. Since $\tau$ has constant complexity, $\Phi_\tau$ is a 
semi-algebraic surface of constant complexity. A point $\ell^* \in L^*$ lies in $\Phi_\tau$ if and only if 
there is a surface $\gamma_\sigma$, with $\sigma^*\in\tau$, that passes 
below $\ell^*$. This happens when, back in $\Sspace$, the surface $\omega_\ell$ (corresponding to the line $\ell$) crosses $\tau$ or lies below $\tau$.
By construction, there are at most 
$n/r + n/r = 2n/r$ such surfaces. Consequently, $\Phi_\tau$ is $(2n/r)$-shallow with respect to the points 
of $L^*$. 

Set $\Phi = \{\Phi_\tau \mid \tau \in \Xi\}$.
$\Phi$ is a family of $O^*(r^4)$ constant-complexity semi-algebraic surfaces\footnote{%
  By construction, as in~\cite{ShSh}, these semi-algebraic sets do not correspond to spheres any more, 
  but they are nevertheless semi-algebraic sets of constant complexity.} 
in $\Lspace$, each of which is $(2n/r)$-shallow with respect to $L^*$. This is our desired test set, as stated in the following lemma.
The proof of the lemma is an immediate consequence of our construction.
\begin{lemma} \label{lem:btest}
	Let $\sigma$ be a sphere that satisfies (C1) and (C2) with respect to the lines of $L$ and that 
	is $(n/r)$-shallow with respect to $L$.  Then there exists a semi-algebraic set of $\Phi$ that 
	contains $\sigma^*$.
\end{lemma}
Plugging Lemma~\ref{lem:btest} into Lemma~\ref{lem:ms}, $\Phi$ is a test set for $L^*$ with respect 
to the semi-algebraic ranges in $\Gamma$, with $\delta=1$ and $\zeta(m) = O^*(m^3)$. 
The bound on $\zeta(m)$ follows from Theorem~\ref{thm:env4d}.
Putting everything together, we thus obtain:

\begin{theorem}
	\label{thm:NNPL}
  A set $L$ of $n$ lines in $\reals^3$ can be preprocessed,
	in $O^*(n)$ expected time, into a data structure of size $O(n)$
	so that for any query point $p\in\reals^3$, the line of $L$ 
  nearest to $p$ can be computed in $O^*(n^{2/3})$ time. 
\end{theorem}



\section{Nearest-Neighbor Queries with Lines in $\reals^3$}
\label{sec:Fast_NN}


In this section we consider the converse situation, where queries are lines in $\reals^3$. We first consider in Section~\ref{sec:Fast_NN1} a simpler, yet challenging, case where
the input is a set of points in $\reals^3$, and then, in Section~\ref{sec:Fast_NN2}, consider 
the case where the input is a set of lines in $\reals^3$. We 
are interested in a data structure that answers NN queries in $O^*(1)$ time using
as little storage as possible.

\subsection{Nearest-point queries with lines in $\reals^3$}
\label{sec:Fast_NN1}

Let $P$ be a set of $n$ points in $\reals^3$. Since we are aiming for an $O^*(1)$ query time, we work in 
the $4$-dimensional parametric space $\Lspace$ of (query) lines (the same parametric space used in the previous section), where
a line $\ell$ in $\reals^3$, given by the equations $y=ax+c$ and $z=bx+d$, is represented as the 
point $\ell^* = (a,b,c,d)\in \Lspace$. We begin by describing the distance function between a point and a line in $\reals^3$ and the Voronoi diagram that the points of $P$ induce in $\Lspace$.

\paragraph{Distance function, lower envelope, Voronoi diagram.}
Let $\ell^* = (a,b,c,d) \in \Lspace$.
For a fixed pair $a, b\in\reals$, the (unnormalized) direction of $\ell$, $(1,a,b)$,
is fixed. Let $H$ be the plane that is orthogonal to $\ell$ (i.e., with normal direction $(1,a,b)$) 
and passes through the origin.
Redefine the representation of $\ell$ so that $(c,d)$ is actually the intersection 
of $\ell$ with $H$, in a suitable canonical coordinate frame within $H$ (we omit here the easy details of specifying this frame, noting that it does depend on $(a,b)$). Write $u=(1,a,b)$.

For a point $p\in P$, let $p^\downarrow$ denote its projection onto $H$. Concretely, write $p^\downarrow = p + tu$. 
The condition for $p^\downarrow$ to lie in $H$ is that $p+tu$ be orthogonal to $u$ (recall that $H$ passes 
through the origin). That is, we require that
\[
(p+tu)\cdot u = p\cdot u + t|u|^2 = 0,\quad\text{or}\quad 
t = -\frac{p\cdot u}{|u|^2} .
\]
That is, we have
\[
p^\downarrow = p - \frac{p\cdot u}{|u|^2} u .
\]
Write $p^\downarrow = (x_p(a,b),y_p(a,b))$; clearly, these coordinates depend on $(a,b)$.
The distance between $p$ and $\ell$, denoted by $\dist(p,\ell)$, is then the distance between 
$p^\downarrow$ and $(c,d)$. That is,
\begin{align}
	\label{eq:dist}
\dist^2(p,\ell) & = (x_p(a,b) - c)^2 + (y_p(a,b) - d)^2 \nonumber \\
& = (x_p^2(a,b) + y_p^2(a,b)) - 2cx_p(a,b) - 2dy_p(a,b) + (c^2+d^2) .
\end{align}
For a query line $\ell$, our goal is to compute $\arg\min_{p\in P} \dist^2(p,\ell)$,
the point $p\in P$ that is closest to $\ell$, i.e., minimizes (\ref{eq:dist}).
Since $c^2+d^2$ is common to all points $p$, we can drop it, and seek the point $p$ that minimizes
\begin{equation} \label{eq:fp}
f_p(a,b,c,d) = g_p(a,b) - 2cx_p(a,b) - 2dy_p(a,b) ,
\end{equation}
where $g_p(a,b) = x_p^2(a,b) + y_p^2(a,b)$.
Let $\F = \{f_p \mid p\in P\}$ be the resulting set of
$n$ $4$-variate functions.  Consider the lower envelope $E: \Lspace \rightarrow \reals$ of $\F$ defined as
\[
E(a,b,c,d) = \min_{p\in P} f_p(a,b,c,d).
\]
The projection of the graph of $E$ onto $\Lspace$, denoted by $\M := \M(P)$, is called the minimization diagram 
of $\F$. $\M$ induces a partition of $\Lspace$, to which we refer as the Voronoi diagram of $P$ in $\Lspace$.
Each cell $\tau$ of $\M$ is associated with
a point $p \in P$ that is the nearest neighbor of all lines whose dual points lie in the cell $\tau$. For a query 
line $\ell$, we wish to 
locate the cell of $\M$ containing $\ell^* = (a,b,c,d)$. However, currently we do not know how to preprocess 
four-dimensional minimization diagrams, like $\M$, into a data structure of size $O^*(n^4)$ for answering point-location queries in $O^*(1)$ time. We manage
to address this problem by exploiting the additional structure of the Voronoi cells of $\M$.

\paragraph{Structure of Voronoi cells.}
For each point $p\in P$, let $\M_p$ denote the region of $\Lspace$ where $f_p$
attains $E$, i.e., the set of cells of $\M$ that are associated with $p$. Let $E_p$ denote the graph of $E$ restricted to $\M_p$, which is a suitable subset of the
graph of $f_p$.

\begin{figure}[htb]
   \begin{center}
     {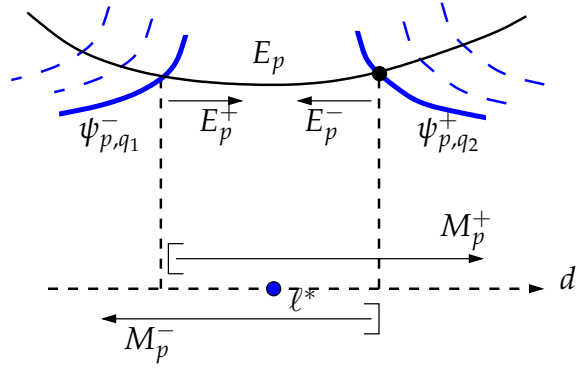 } 
   \end{center}
   \caption{
     The structure of the decomposition of the lower envelope and the minimization diagram of the sample.
     To simplify the figure, the superscripts $(d+)$ and $(d-)$ have been suppressed.}
   \label{fig:EMpm}
\end{figure}

For each $q\in P$, $q\ne p$, let $\sigma_{p,q}$ denote the intersection surface of $f_p$ and $f_q$, 
which is a three-dimensional surface that is disjoint from the relative interior of $E_p$, and does not pass below any point on $E_p$. 
It is defined by the equation
\[
g_p(a,b) - 2cx_p(a,b) - 2dy_p(a,b) = g_q(a,b) - 2cx_q(a,b) - 2dy_q(a,b).
\]
Assuming $y_q(a,b) \ne y_p(a,b)$, we define a trivariate function $\psi_{p,q}: \Lspace^{(d)} \rightarrow \reals$, where $\Lspace^{(d)} \subset \Lspace$ is the $3$-dimensional hyperplane $d=0$, as follows:
\begin{equation} \label{eq:psipq}
d = \psi_{p,q}(a,b,c) := \frac{g_q(a,b) - g_p(a,b)}{2(y_q(a,b) - y_p(a,b))} - \frac{x_q(a,b) - x_p(a,b)}{y_q(a,b) - y_p(a,b)} c .
\end{equation}
The surface $\sigma_{p,q}$ partitions $\Lspace$ into the two regions
\[
	K_{p,q}^{(d+)}  = \{ \ell^*\in\Lspace \mid f_p(\ell^*) \le f_q(\ell^*) \} \quad\mbox{and}\quad
	K_{p,q}^{(d-)}  = \{ \ell^*\in\Lspace \mid f_p(\ell^*) \ge f_q(\ell^*) \} .
\]
Then $\M_p=\bigcap_{q\in P\setminus\{p\}} K_{p,q}^+$. By (\ref{eq:psipq}), 
we can write $K_{p,q}^{(d+)}$ as
\begin{align*}
  K_{p,q}^{(d+)} & = 
  \{ (a,b,c,d) \in \Lspace \mid y_q(a,b) - y_p(a,b) \ge 0, \; d \le \psi_{p,q}(a,b,c) \} \bigcup \\
  & \{ (a,b,c,d)\in \Lspace  \mid y_q(a,b) - y_p(a,b) \le 0, \; d \ge \psi_{p,q}(a,b,c) \} .
\end{align*}
To simplify this representation, we define two functions $\psi_{p,q}^+, \psi_{p,q}^- : \Lspace^{(d)} \rightarrow \reals$ by:
\begin{align}
  \label{eq:surfaces}
  \begin{split}
    \psi_{p,q}^{(d+)}(a,b,c) & = 
    \begin{cases}
      \psi_{p,q}(a,b,c) & y_q(a,b) - y_p(a,b) \ge 0 \\
      +\infty & \text{otherwise}
    \end{cases} \\
    \psi_{p,q}^{(d-)}(a,b,c) & = 
    \begin{cases}
      \psi_{p,q}(a,b,c) & y_q(a,b) - y_p(a,b) \le 0 \\
      -\infty & \text{otherwise} .
    \end{cases} 
  \end{split}
\end{align}
Then we can write
\begin{equation} \label{eq:kpq}
  K_{p,q}^{(d+)} = \{ (a,b,c,d) \in \Lspace \mid 
  \psi_{p,q}^{(d-)}(a,b,c) \le d \le \psi_{p,q}^{(d+)}(a,b,c) \} . 
\end{equation}
In other words, $\M_p$ is the \emph{sandwich region} between the lower envelope (with respect to the $d$-direction)
$E_p^{(d-)}$ of the functions $\psi_{p,q}^{(d+)}$ and the upper envelope $E_p^{(d+)}$ of the functions 
$\psi_{p,q}^{(d-)}$, for $q\in P\setminus \{p\}$. See Figure~\ref{fig:EMpm} for an illustration.

We can thus write $\M_p$ as $\M_p^{(d-)}\cap \M_p^{(d+)}$, where $\M_p^{(d-)}$ (resp., $\M_p^{(d+)}$) is the region below 
the lower envelope $E_p^{(d-)}$ (resp., above the upper envelope $E_p^{(d+)}$) in the $d$-direction. 

Note that the above construction is symmetric in $c$ and $d$, as each function $f_p$ is linear 
in both $c$ and $d$. We can therefore repeat the whole construction, switching between $c$ and $d$.
The analysis is fully symmetric, with obvious modifications, such as having $x_q(a,b)-x_p(a,b)$ in the 
denominators in (\ref{eq:fp}), and similar straightforward changes.  $\M_p$ can now be written as 
$\M_p^{(c-)}\cap \M_p^{(c+)}$, where $\M_p^{(c-)}$ (resp., $\M_p^{(c+)}$) is the region below (resp., above),
in the $c$-direction, the lower envelope $E_p^{(c-)}$ (resp., upper envelope $E_p^{(c+)}$) of the corresponding set of 
trivariate functions $\psi_{p,q}^{(c-)}$ (resp., $\psi_{p,q}^{(c+)}$) defined analogously to 
$\psi_{p,q}^{(d-)}$ (resp., $\psi_{p,q}^{(d+)}$).

We conclude this discussion with the following observation, which will be the key to the performance of our data structure and the query procedure. 

\begin{lemma}
  \label{lem:nearer-point}
  Let $\ell^* = (\ell_a,\ell_b,\ell_c,\ell_d)\in \Lspace$, and let $p$ be a point of $P$. Let
  $\rho^{(d)}$ (resp., $\rho^{(c)}$) denote the line in the $d$-direction (resp., $c$-direction) 
  in $\Lspace$ passing through $\ell^*$, 
  and let $\gamma^{(d)}$ (resp., $\gamma^{(c)}$) denote the curve on (the graph of) $f_p$ traced over 
  the line $\rho^{(d)}$ (resp., $\rho^{(c)}$). 
  Let $q$ be a point of $P$ that is nearer to $\ell$ than $p$, assuming that such a point exists,
  i.e., $f_q(\ell^*) < f_p(\ell^*)$. Then 
  $f_q$ intersects either $\gamma^{(d)}$ or $\gamma^{(c)}$. Furthermore if 
  $f_q$ intersects $\gamma^{(d)}$ at a point $w=(w_a,w_b,w_c,w_d)$ such that 
  $w_d > \ell_d$ (resp., $w_d < \ell_d$) then we have 
	$w_d = \psi_{p,q}^{(d-)}(w_a,w_b,w_c)$  (resp., $w_d=\psi_{p,q}^{(d+)}(w_a,w_b,w_c)$). A similar 
  property holds if $f_q$ intersects $\gamma^{(c)}$.
\end{lemma}

\noindent\textbf{Proof.}
  Suppose $f_q$ does not intersect $\gamma^{(d)}$. Then we would have, using (\ref{eq:fp}),
  \[
  g_q(a,b) - 2cx_q(a,b) - 2dy_q(a,b) < g_p(a,b) - 2cx_p(a,b) - 2dy_p(a,b)
  \]
  for every $d$. Since $a,b,c$ are fixed along $\gamma^{(d)}$, this can happen only when 
  $y_q(a,b) = y_p(a,b)$.  Repeating the same argument for $\gamma^{(c)}$, if $f_q$ does not intersect $\gamma^{(c)}$,
  then $x_q(a,b) = x_p(a,b)$. Therefore, if $f_q$ does not intersect either of these curves then we also have, by
  definition,	$g_q(a,b) = g_p(a,b)$, which implies that $f_q(\ell^*) = f_p(\ell^*)$, i.e., $p$ and $q$ are 
  equidistant from $\ell$. This contradicts the assumption that $q$ is (strictly) nearer to $\ell$ than $p$.
  
  Thus $f_q$ intersects one of the curves, say, for specificity, that it intersects $\gamma^{(d)}$. 
  Again, by (\ref{eq:fp}), $f_q$ intersects $\gamma^{(d)}$ at a unique point $w=(w_a,w_b,w_c,w_d)$, 
  with $w_d=\psi_{p,q}(w_a,w_b,w_c)$. If $w_d>\ell_d$ (resp., $w_d<\ell_d$), then by (\ref{eq:surfaces}), 
	we must have $w_d = \psi_{p,q}^{(d-)}(w_a,w_b,w_c)$ (resp., $w_d=\psi^{(d+)}(w_a,w_b,w_c)$). This completes the proof of the lemma.
  $\Box$

We are now ready to describe the data structure based on the above lemma.



\paragraph{Overall data structure.}
Fix some sufficiently large constant parameter $r>0$.  We choose a random subset
$R \subseteq P$ of $cr\log r$ points, for a suitable absolute constant $c>0$.
We construct the Voronoi diagram $\M(R)$ of $R$. 
For every point $p\in R$, we construct $\M_p^{(d-)}, \M_p^{(d+)}, \M_p^{(c-)}, \M_p^{(c+)}$,
as defined above (with respect to $\M(R)$).
Let $\Xi_p^{(d-)} = \VD(\M_p^{(d-)})$ be the vertical decomposition of $\M_p^{(d-)}$. 
Similarly define, $\Xi_p^{(d+)}, \Xi_p^{(c-)}, \Xi_p^{(c+)}$. 
Let $\Xi$ be the set of cells in all these $4|R|$ vertical decompositions.
By Theorem~\ref{thm:env4d}, $|\Xi|= O^*(r\cdot r^3) = O^*(r^4)$, and
by Theorem~\ref{thm:env_algo}, $\Xi$ can be constructed in a total of $O^*(r^4)$ 
expected time. 

We define a \emph{conflict list} $L_\tau$ for every $\tau\in\Xi$, as follows.
For each point $p\in R$ and each cell $\tau$ of $\Xi_p^{(d-)}$ (resp., $\Xi_p^{(d+)}$), we define 
$P_\tau \subset P$ to be the subset of points $q\in P$ for which the surface 
$d = \psi_{p,q}^{(d+)}$ (resp., $d= \psi_{p,q}^{(d^-)}$) crosses $\tau$. With a suitable choice of $c$,
the size of each conflict list is at most $n/r$, with high probability, because, by construction, for a cell $\tau$ of $\M_p^{(d-)}$ (resp., $\M_p^{(d+)}$), none of the 
surfaces $d=\psi_{p,u}^+$ (resp., $d=\psi_{p,u}^-$), for $u\in R\setminus \{p\}$, intersect $\tau$~\cite{HW87}.
Similarly we define the conflict lists of cells in $\Xi_p^{(c-)}, \Xi_p^{(c+)}$; their sizes are also all at most $n/r$, with high probability. 

For each cell $\tau\in\Xi$, 
we recursively build the data structure on $P_\tau$. The recursion stops when the size of a subproblem becomes smaller than some fixed absolute constant $n_0$.
Since there are $O^*(r^4)$ subproblems of size at most $n/r$ each, 
a straightforward analysis shows that the size of the  overall structure is $O^*(n^4)$, and that
it can be constructed in $O^*(n^4)$ expected time.

\paragraph{Query procedure.}
A query with a line $\ell$ is processed as follows. We compute the nearest neighbor of $\ell$ in $R$, which we call $p$. Next, we compute the cells
$\tau^{(d-)}, \tau^{(d+)}, \tau^{(c-)}, \tau^{(c+)}$ of 
$\M_p^{(d-)}, \M_p^{(d+)}, \M_p^{(c-)}, \M_p^{c+)}$, respectively, that contain $\ell^*$. All this is
done in brute force and takes $O^*(1)$ time. If $P$ contains a point $q$ that is nearer to $\ell$ than $p$, then by
Lemma~\ref{lem:nearer-point}, $f_q$ intersects either the curve $\gamma^{(d)}$ or $\gamma^{(c)}$. 
Suppose $f_q$ intersects $\gamma^{(d)}$ at a point $w=(w_a,w_b,w_c,w_d)$.
Again, by Lemma~\ref{lem:nearer-point}, if $w_d\ge \ell_d$,
then $w_d = \psi_{p,q}^-(w_a,w_b,w_c)$, implying that  
$w \in \tau^{d+}$ and thus $q$ belongs to the conflict list $P_{\tau^{d+}}$.
Similarly, if $w_d<\ell_d$, then $q$ belongs to the 
conflict list $P_{\tau^{d-}}$. A symmetric analysis applies when $f_q$ intersects $\gamma^{(c)}$. 
In summary, if $q$ is closer to $\ell$ than $p$ then $q$ lies in the conflict lists of one of
$\tau^{(d-)}, \tau^{(d+)}, \tau^{(c-)}, \tau^{(c+)}$. Hence, we need to search recursively in these four 
subproblems, and return the nearest point among $p$ and the points returned by these four recursive subproblems.

Since we recurse in four subproblems, each of size at most $n/r$ (and $r$ can be chosen to be a 
sufficiently large constant), the total query time is $O^*(1)$ (it is not polylogarithmic, though). We thus obtain the following result:

\begin{theorem} \label{thm:n4}
  A given set $P$ of $n$ points in $\reals^3$ can be preprocessed,
	in $O^*(n^4)$ expected time, into a data structure of size $O^*(n^4)$,
	so that, for any query line $\ell\in\reals^3$, the point of $P$ 
  nearest to $\ell$ can be computed in $O^*(1)$ time. 
\end{theorem}

\subsection{Nearest-line queries with lines in $\reals^3$}
\label{sec:Fast_NN2}


Next, we show that the machinery in the preceding subsection can be extended 
(with a couple of twists---see below) to obtain a line NN-searching data structure, with the same asymptotics
performance, when the input is a set $L$ of $n$ lines in $\reals^3$, and we want to find the line nearest to a query line.
We first describe the two new challenges we face in dealing with lines as input, and explain how to address them, and then 
describe the overall data structure.

We use the same representation $(a,b,c,d)$ for the query line $\ell$, using the orthogonal plane $H$ as 
before.  Thus $\ell$ is represented as the same point $\ell^*\in\Lspace$.
For a line $\lambda\in L$, 
let $\lambda^\downarrow$ denote the projection of $\lambda$ onto $H$. A crucial 
observation, which is easy to verify, is that
\[
	f_\lambda(\ell^*) := \dist(\ell,\lambda) = \dist(\ell,\lambda^\downarrow) = \dist((c,d),\lambda^\downarrow) .
\]
The equation of $\lambda^\downarrow$, in the canonical coordinate frame within $H$, is of the form
\[
\xi_\lambda(a,b) x + \eta_\lambda(a,b) y + \zeta_\lambda(a,b) = 0 ,
\]
where we normalize the coefficients so that
$\xi_\lambda^2(a,b) + \eta_\lambda^2(a,b) = 1$. Hence,
\begin{equation}
	\label{eq:l-dist}
	f_\lambda(a,b,c,d)  = \left| \xi_\lambda(a,b) c + \eta_\lambda(a,b) d + \zeta_\lambda(a,b) \right| .
\end{equation}
Except for the absolute value, (\ref{eq:l-dist}) is linear in $c$ and $d$, as in the preceding analysis,
a property that has been crucial for the analysis there, and will be crucial for the analysis here too. 

We handle the absolute value as follows.
Orient each line $\lambda\in L$ in an arbitrary (but fixed) manner, say in the positive $x$-direction, and similarly orient each query line $\ell$.
If we know the relative orientation of $\ell$ and $\lambda$,
then we also know the sign in the expression for $f_\lambda(\ell^*)$.
In fact, we can reduce the setup in such a way that allows us to assume that 
the sign is positive if and only if the relative orientation is positive.
For a line $\lambda\in L$, we define the surface $\sigma_\lambda\subset\Lspace$, which is the locus of all 
points $\ell^*\in\Lspace$ such that $\ell$ touches $\lambda$.
It partitions $\Lspace$ space into two portions, one consisting of points representing lines that 
are positively oriented with respect to $\lambda$, and the other consists of points with 
negative orientations. 
We construct a data structure on these surfaces that, for a query (oriented) line $\ell$, partitions the set of all lines of $L$ into $O(\log n)$ ``canonical'' subsets such that, for every canonical subset, either all its lines are positively oriented with respect to $\ell$ or all of them are negatively oriented.

In view of the above discussion, let us assume that the query line has positive orientation with respect
to all lines in $L$, and that this corresponds to a positive sign of the expression in (\ref{eq:l-dist}). 
We construct a data structure on $L$ using, more or less, the same 
machinery as in Section~\ref{sec:Fast_NN1}, exploiting the double linearity (in $c$ and $d$) of the distance functions. Here we face the second challenge.
Recall that we basically showed in Lemma~\ref{lem:nearer-point} 
that if $f_p$ and $f_q$ do not cross along the lines $\rho_d, \rho_c$, then we have $x_p(a,b) = x_q(a,b)$ and $y_p(a,b) = y_q(a,b)$, and thus the free terms $g_p(a,b)$ and $g_q(a,b)$ are also equal, implying that $p$ and $q$ are equidistant 
from the query line $\ell$.
Here, in contrast, if $f_\lambda, f_{\lambda'}$, for two distinct lines $\lambda, \lambda' \in L$, do not cross 
along $\rho_c, \rho_d$, we can show, using the same reasoning as before, but based  on  (\ref{eq:l-dist}), that 
$\xi_\lambda(a,b) = \xi_{\lambda'}(a,b)$ and $\eta_\lambda(a,b) = \eta_{\lambda'}(a,b)$ (actually, one equality suffices, because of our normalization). However, now it no longer follows that 
$\zeta_\lambda(a,b) = \zeta_{\lambda'}(a,b)$. That is, the projected lines (on $H(a,b)$) could be parallel, 
and $\lambda'$ could still be (strictly) nearer to $\ell$ than $\lambda$. 

To address this issue we proceed as follows. For each pair of lines $\lambda$, $\lambda'$ in $L$, let 
$\beta_{\lambda,\lambda'}$ denote the one-dimensional locus of all $(a,b)$ for which the projections of 
$\lambda$ and $\lambda'$ onto $H$ are parallel; this is the curve $\xi_\lambda(a,b) = \xi_{\lambda'}(a,b)$. 
For each $\lambda$ in the sample $R$, we construct the two-dimensional arrangement $\A_\lambda$ of the 
curves in $\{\beta_{\lambda,\lambda'}\mid \lambda'\in L\setminus \{\lambda\}\}$, 
in the $(a,b)$-plane. For a query dual point
$\ell^*=(\ell_a, \ell_b, \ell_c, \ell_d)$, 
we locate the point $(\ell_a,\ell_b)$ in $\A_\lambda$ and
find the set $L_{\rm par}$ of the curves $\beta_{\lambda,\lambda'}$ that contain the point $(\ell_a,\ell_b)$
to determine the lines of $L$ whose projections onto $H(a,b)$ are parallel to $\lambda$. 
(See below how the algorithm handles sets $L_{\rm par}$ of large size.)

We now describe the overall data structure and the query procedure by incorporating these observations in the 
data structure described in Section~\ref{sec:Fast_NN1}.


\paragraph{Overall data structure.}
We build a three-level data structure. Let $\Sigma=\{\sigma_\lambda\mid\lambda \in L\}$. 
At the top-level, we construct a tree data structure $\T^{(1)}$ for answering point-enclosure queries on $\Sigma$,
using the algorithm in~\cite{AAEZ}. Each  node $u$ of $\T^{(1)}$ is associated with a \emph{canonical subset} 
$L_u\subseteq L$ of lines. For a query line $\ell$, querying with $\ell^*$ in $\T^{(1)}$ 
partitions the lines of $L$ into $O(\log n)$ canonical subsets, each associated with one of its nodes,
such that all lines in one subset are either positively oriented with respect to $\ell$ 
or all of them are negatively oriented. 

For each node $v$ of $\T^{(1)}$, we construct two second-level data 
structures $\T^{(2+)}_v, \T^{(2-)}_v$ on the canonical subset $L_v$---one assuming that the sign in (\ref{eq:l-dist})
is positive and the other assuming that it is negative. These structures are constructed by following and adapting 
the construction in Section~\ref{sec:Fast_NN1}, using the expressions in (\ref{eq:l-dist}) (without the
absolute value) instead of those in (\ref{eq:fp}), following both the $c$- and $d$-directions, and using 
partial lower envelopes within the minimization diagram.
Each of $\T^{(2+)}_v, \T^{(2-)}_v$ essentially consists of several tree data structures. Each node $w$ of $\T^{(2+)}$ or $\T^{(2-)}$ is also associated with a 
subset $L_w \subseteq L_v$ of lines. We choose a random subset $R_w \subset L_w$ of size $cr\log r$, for some constant $c\ge 1$, and 
construct, as in Section~\ref{sec:Fast_NN1}, a total of $O^*(r^4)$ subproblems, each of size at most $|L_w|/r$. In addition, we 
now store the following third-level structure at $w$: For each line $\lambda \in R$, we construct 
the two-dimensional arrangement $\A_\lambda$ of the curves 
$\B_\lambda = \{\beta_{\lambda,\lambda'} \mid \lambda' \in L_w\setminus R_w\}$ and preprocess it for 
point-location queries. 
If the input lines are in general position, then at most two curves of $\B_\lambda$ 
pass through any point $(a,b)$, and 
we simply store them. Otherwise, many curves of $\B_\lambda$ may pass through a vertex $\chi=(\chi_a,\chi_b)$ of 
$\A_\lambda$. Let $L_\chi \subseteq L_w\setminus R_w$ be the subset of lines whose curves are incident 
on $\chi$. We store $L_\chi$ in a sorted order (by the ordering of their projections on the plane 
$H(\chi_a,\chi_b)$) so that for a query line $\ell$ of the form $\ell^*=(\chi_a,\chi_b,\ell_c,\ell_d)$, we can find 
the line in $L_\chi$ nearest to $\ell$ in $O(\log n)$ time. The total size of this third-level data 
structure over all lines of $R$ is $|R| \cdot O(|L_w|^2)= O(|L_w|^2)$.
Using the properties of multi-level data structures, one can show that the overall size of the data structure is $O^*(n^4)$ and that it can be constructed in $O^*(n^4)$ expected time.


\paragraph{Query procedure.}

For a query line $\ell$, we first search in $\T^{(1)}$ with $\ell^*$ and compute a partition of $L$ into
$O(\log n)$ canonical subsets, each associated with a node of $\T^{(1)}$, such that each subset is positively
oriented or negatively oriented with respect to $\ell$.
For each such node $v$, if the lines in $L_v$ have positive (resp., negative) orientation with respect to $\ell$, 
we search in $\T^{(2+)}_v$ (resp.\ $\T^{(2-)}_v$) with $\ell^*$, as in Section~\ref{sec:Fast_NN1}. 
At each second-level node $w$ visited by the query procedure, if $\lambda$ is the nearest neighbor of 
$\ell$ in $R_w$, we recursively search in the four corresponding children of $v$ as in the previous section.
In addition, we locate the point $(\ell_a^*, \ell_b^*)$ in the arrangement
$\A_\lambda$ to find, in $O(\log n)$ time, the nearest neighbor of $\ell$ among the 
line of $L_w\setminus R$ whose projections on $H(\ell_a,\ell_b)$ are parallel to that of $\lambda$, if any such lines exist.
Following the same analysis as above, the overall query time remains $O^*(1)$.
Putting everything together, we obtain the following result:
\begin{theorem} \label{thm:n4lines}
  A given set $L$ of $n$ lines in $\reals^3$ can be preprocessed, in $O^*(n^4)$ expected time, 
	into a data structure of size $O^*(n^4)$, so that, for any query line $\ell\in\reals^3$, the 
	line of $L$ nearest to $\ell$ can be computed in $O^*(1)$ time. 
\end{theorem}

\section{All Line-Point Nearest Neighbors in $\reals^3$}
\label{sec:ann}


Here we consider an 
offline version of the problem studied in the previous sections.

\subsection{A simple offline algorithm}
\label{sec:ann1}

Let $L$ be a set of $n$ lines and $P$ a set of $m$ points in $\reals^3$. Our goal is to
compute, for each line $\ell\in L$, the point of $P$ that is nearest to $\ell$. This is the batched, or offline, 
version of the line-point nearest-neighbor problem studied in Section~\ref{sec:Fast_NN}.

We first present a rather simple algorithm, which we will improve in the next subsection, using
a more involved analysis.

Our approach consists of the following steps. We first take a random sample $R$ of $t$ points of
$P$, for a parameter $t$ that we will set later. 
For each line $\ell\in L$ we compute the point $p\in R$ that is nearest to $\ell$, and
associate with $\ell$ the cylinder $C_\ell$ that has $\ell$ as its symmetry axis, and has
radius $\dist(p,\ell)$. The overall cost of this step is $O(nt)$, using a brute-force approach.

By standard random sampling arguments, $C_\ell$ contains at most
$O\left(\frac{m}{t}\log t\right) = O^*(m/t)$ points of $P$, which holds, with high probability, 
for all lines $\ell\in L$. Let $\C$ denote the set of these $n$ cylinders, and denote by $K = O^*(mn/t)$ the total number of point-cylinder containments.

We next perform an offline point-enclosure reporting step, 
where the cylinders of $\C$ are our input, each query is with a point $p\in P$, and the goal 
of the query is to report all the cylinders that contain $p$. We do this using the algorithm presented in Section~\ref{sec:point_enclosure_cyl}. We apply this step to each point 
of $P$. Each line $\ell\in L$ collects the points $p\in P$ for which $C_\ell$ contains $p$,
and outputs the nearest point to $\ell$. By the analysis in Section~\ref{sec:point_enclosure_cyl},
the point enclosure queries take a total of
$O^*(n^2 + m + K) = O^*(n^2 + m + mn/t)$ time. Including the cost of the sampling as described above,
the overall cost is $O^*(n^2 + m + mn/t + nt)$, which we optimize by choosing $t = m^{1/2}$. The resulting bound,
$O^*(n^2 + m + m^{1/2}n)$, can be trivially improved by breaking the set of lines into subsets, 
each of size at most $m^{1/2}$, and by repeating the above procedure to each subset and all the points. 
The resulting running time is $O^*(m^{1/2}n + m)$. That is, we have:
\begin{proposition} \label{thm:annx}
Given $m$ points and $n$ lines in $\reals^3$, we can compute, for each line $\ell$, the point nearest to
$\ell$ in $O^*(m^{1/2}n + m)$ overall randomized expected  time.
\end{proposition}

\subsection{An improved algorithm}
\label{sec:ann_imp}


We next present an improved and faster algorithm.
The improvement is in the implementation of the point enclosure procedure amid the cylinders of $\C$. 
It is obtained by combining the machinery of the algorithm in Section~\ref{sec:Fast_NN} with the point enclosure 
mechanism of Section~\ref{sec:point_enclosure_cyl}, and proceeds as follows.

We run a modified version of the procedure of Section~\ref{sec:point_enclosure_cyl}. 
At each recursive step, we obtain a decomposition of the problem into $O^*(r^4)$ subproblems, for the 
constant parameter $r$ used there, each involving at most $m/r$ points of $P$ (where $m$ is the size 
of the current point set). The lines of $L$ are represented as $n$ points in the four-dimensional line space $\Lspace$ (where $n$ is the size 
of the current line set). By the analysis in Section~\ref{sec:Fast_NN}, adapted to the offline setup,
each line participates in at most four of the subproblems. From this it easily follows that we can split each
subproblem into further subproblems, so that the number of subproblems remains $O^*(r^4)$, and each subproblem
involves at most $m/r$ points of $P$ and at most $n/r^4$ lines of $L$. Moreover, by construction, if 
$\ell\in L$ and $p$ is the point of $P$ nearest to $\ell$ then at least one of the subproblems involves both $\ell$ and $p$.

We carry the recursion for $j$ levels, for some parameter $j$ that we will fix shortly. At the bottom of this
prematurely terminated recursion, we have $O^*(r^{4j})$ subproblems, each involving at most $m/r^j$ points
of $P$ and at most $n/r^{4j}$ lines of $L$.

We now apply to each subproblem the simpler algorithm in 
Section~\ref{sec:ann1}. By Proposition~\ref{thm:annx}, 
this costs a total of
\[
O^*(r^{4j}) \cdot O^*\left(
\left( \frac{m}{r^j} \right)^{1/2} \frac{n}{r^{4j}} 
+ \frac{m}{r^j} \right) =
O^*\left( \frac{m^{1/2}n}{r^{j/2}} + mr^{3j} \right) .
\]
We now set $j$ so as to roughly balance these terms, i.e., choose $j$ to satisfy $r^{7j/2} = n/m^{1/2}$,
or $r^j = n^{2/7}/m^{1/7}$. Substituting this in the above bound, we obtain the overall cost $O^*(m^{4/7}n^{6/7})$.
For this to make sense, we require that $m/r^j \ge 1$ and $n/r^{4j} \ge 1$, and that $r^j\ge 1$.
As is easily checked, this means that this choice of $j$ makes sense when $n^{1/4}\le m \le n^2$.
When $m < n^{1/4}$, we only apply the procedure of Section~\ref{sec:Fast_NN1}, which takes a total of
$O^*(m^4 + n) = O^*(n)$ time.
When $m > n^2$, we only apply the procedure of Section~\ref{sec:ann1}, which takes a total of
$O^*(m^{1/2}n + m) = O^*(m)$ time. Altogether we obtain

\begin{theorem} \label{thm:ann_imp}
  Given sets $P$ of $m$ points and $L$ of $n$ lines in $\reals^3$, we can compute, for each line 
  $\ell\in L$, the point of $P$ nearest to $\ell$, in overall $O^*(m^{4/7}n^{6/7} + m + n)$ randomized expected time.
\end{theorem}
\section{Conclusion}
In this paper, we settled in the affirmative a few long-standing open problems involving the 
vertical decomposition of various substructures of arrangements in $d=3,4$ dimensions. In particular, we obtained
sharp bounds on the vertical decomposition of the complement of the union of a family of 
semi-algebraic sets in $\reals^3$ of constant complexity, and of the lower envelope of a family of semi-algebraic 
trivariate functions of constant complexity. We also obtained an output-sensitive bound on the size of the 
vertical decomposition of the full arrangement of a family of semi-algebraic sets in $\reals^3$
of constant complexity. These results lead to 
efficient algorithms for constructing the vertical decompositions themselves, for constructing 
$(1/r)$-cuttings of the above substructures of arrangements, and for answering point-enclosure queries. 
Finally, we applied these results to obtain faster data structures for various basic proximity problems 
involving lines and points in $\reals^3$. 

We conclude by mentioning a few open problems:
\begin{itemize}
\item The major open question is, of course, to improve the complexity of the vertical 
  decomposition of the arrangement of a family of semi-algebraic sets in $\reals^d$ 
  for $d\ge 5$. But an immediate open question is whether the techniques developed in this 
  paper can be extended to obtain improved bounds on the vertical decomposition of 
  various substructures of arrangements (besides lower or upper envelopes) in $\reals^4$.
\item No non-trivial lower bounds are known for nearest-neighbor data structures
  involving lines in $\reals^3$. This raises the question whether
  the data structures presented in Sections~\ref{sec:plnn} and~\ref{sec:Fast_NN} 
  are (almost) best possible, or whether one can obtain significantly faster data structures. 
  For example, can the nearest neighbor of a line amid a set of points in $\reals^3$ be 
  returned in $O(\log n)$ time using an $O^*(n^3)$ size data structure?
\end{itemize}




\end{document}

%% file: point_line_NN.pdf_t
\begin{picture}(0,0)%
\includegraphics{point_line_NN.pdf}%
\end{picture}%
\setlength{\unitlength}{3315sp}%
\begingroup\makeatletter\ifx\SetFigFont\undefined%
\gdef\SetFigFont#1#2#3#4#5{%
  \reset@font\fontsize{#1}{#2pt}%
  \fontfamily{#3}\fontseries{#4}\fontshape{#5}%
  \selectfont}%
\fi\endgroup%
\begin{picture}(3095,2673)(6819,-5435)
\put(7926,-3819){\makebox(0,0)[lb]{\smash{{\SetFigFont{10}{12.0}{\rmdefault}{\mddefault}{\updefault}{\color[rgb]{0,0,0}$p$}%
}}}}
\put(8831,-3168){\makebox(0,0)[lb]{\smash{{\SetFigFont{10}{12.0}{\rmdefault}{\mddefault}{\updefault}{\color[rgb]{0,0,0}$V_\ell$}%
}}}}
\put(7435,-3730){\makebox(0,0)[lb]{\smash{{\SetFigFont{10}{12.0}{\rmdefault}{\mddefault}{\updefault}{\color[rgb]{0,0,0}$\sigma_p$}%
}}}}
\put(9541,-3436){\makebox(0,0)[lb]{\smash{{\SetFigFont{10}{12.0}{\rmdefault}{\mddefault}{\updefault}{\color[rgb]{0,0,0}$\ell$}%
}}}}
\put(7066,-5371){\makebox(0,0)[lb]{\smash{{\SetFigFont{10}{12.0}{\rmdefault}{\mddefault}{\updefault}{\color[rgb]{0,0,0}$\ell^{-}$}%
}}}}
\put(7111,-4741){\makebox(0,0)[lb]{\smash{{\SetFigFont{10}{12.0}{\rmdefault}{\mddefault}{\updefault}{\color[rgb]{0,0,0}$H_\ell$}%
}}}}
\end{picture}%

%% file: EMpm.pdf_t
\begin{picture}(0,0)%
\includegraphics{EMpm.pdf}%
\end{picture}%
\setlength{\unitlength}{4144sp}%
\begingroup\makeatletter\ifx\SetFigFont\undefined%
\gdef\SetFigFont#1#2#3#4#5{%
  \reset@font\fontsize{#1}{#2pt}%
  \fontfamily{#3}\fontseries{#4}\fontshape{#5}%
  \selectfont}%
\fi\endgroup%
\begin{picture}(3317,2164)(6509,-5440)
\put(9811,-5011){\makebox(0,0)[lb]{\smash{{\SetFigFont{12}{14.4}{\rmdefault}{\mddefault}{\updefault}{\color[rgb]{0,0,0}$d$}%
}}}}
\put(7966,-3661){\makebox(0,0)[lb]{\smash{{\SetFigFont{12}{14.4}{\rmdefault}{\mddefault}{\updefault}{\color[rgb]{0,0,0}$E_p$}%
}}}}
\put(7201,-5371){\makebox(0,0)[lb]{\smash{{\SetFigFont{12}{14.4}{\rmdefault}{\mddefault}{\updefault}{\color[rgb]{0,0,0}$M_p^{-}$}%
}}}}
\put(9091,-4696){\makebox(0,0)[lb]{\smash{{\SetFigFont{12}{14.4}{\rmdefault}{\mddefault}{\updefault}{\color[rgb]{0,0,0}$M_p^{+}$}%
}}}}
\put(6931,-4111){\makebox(0,0)[lb]{\smash{{\SetFigFont{12}{14.4}{\rmdefault}{\mddefault}{\updefault}{\color[rgb]{0,0,0}$\psi_{p,q_1}^{-}$}%
}}}}
\put(8956,-4111){\makebox(0,0)[lb]{\smash{{\SetFigFont{12}{14.4}{\rmdefault}{\mddefault}{\updefault}{\color[rgb]{0,0,0}$\psi_{p,q_2}^{+}$}%
}}}}
\put(7651,-4066){\makebox(0,0)[lb]{\smash{{\SetFigFont{12}{14.4}{\rmdefault}{\mddefault}{\updefault}{\color[rgb]{0,0,0}$E_p^{+}$}%
}}}}
\put(8281,-4066){\makebox(0,0)[lb]{\smash{{\SetFigFont{12}{14.4}{\rmdefault}{\mddefault}{\updefault}{\color[rgb]{0,0,0}$E_p^{-}$}%
}}}}
\put(8191,-5146){\makebox(0,0)[lb]{\smash{{\SetFigFont{12}{14.4}{\rmdefault}{\mddefault}{\updefault}{\color[rgb]{0,0,0}$\ell^*$}%
}}}}
\end{picture}%